\documentclass[a4paper,fleqn,usenatbib,useAMS]{mnras}

\usepackage{graphicx}
\usepackage{txfonts}
\usepackage{color}
\usepackage{url}
\newcommand{\ulyss}{\emph{ULySS}}

\newcommand{\re}{$\rm r_{\rm e}$}

\newcommand{\logg}{$\log g$}

\newcommand{\simlt}{\lower.5ex\hbox{$\; \buildrel < \over \sim \;$}}
\newcommand{\simgt}{\lower.5ex\hbox{$\; \buildrel > \over \sim \;$}}

\newcommand{\kms}{\,km\,s$^{-1}$}

\newcommand{\Msol}{${\rm M}_{\odot}$}
\newcommand{\Rsol}{${\rm R}_{\odot}$}
\newcommand{\Teff}{${\rm T}_{\mbox{\scriptsize eff}}$}
\newcommand{\Feh}{\mbox{$\mbox{[Fe/H]}$}}
\newcommand{\Mh}{\mbox{$\mbox{[M/H]}$}}

\newcommand{\MgFe}{\mbox{$\mbox{[Mg/Fe]}$}}

\newcommand{\alfa}{$\alpha$}

\newcommand{\eg}{\rm{e.g.}}
\def\lesssim{\mathrel{\hbox{\rlap{\hbox{\lower4pt\hbox{$\sim$}}}\hbox{$<$}}}}
\def\gtrsim{\mathrel{\hbox{\rlap{\hbox{\lower4pt\hbox{$\sim$}}}\hbox{$>$}}}}
\def \aj {AJ}
\def \mnras {MNRAS}
\def \apj {ApJ}
\def \apjs {ApJS}
\def \apjl {ApJL}
\def \aap {A\&A}
\def \aaps {A\&AS}
\def \nat {Nature}
\def \araa {ARAA}
\def \pasp {PASP}

\usepackage[T1]{fontenc}
\usepackage{ae,aecompl}

\title[UV-extended E-MILES stellar population models]
{{UV-extended E-MILES stellar population models: 
young components in massive early-type galaxies}}

\author[Vazdekis et~al.]{A. Vazdekis$^{1,2}$\thanks{E-mail:vazdekis@iac.es}, 
M. Koleva$^{3}$, E. Ricciardelli$^{4}$, B. R\"{o}ck$^{1,2}$, J. Falc\'on-Barroso$^{1,2}$\\
$^{1}$Instituto de Astrof\'\i sica de Canarias (IAC), E-38200 La Laguna, Tenerife, Spain\\
$^{2}$Departamento de Astrof\'\i sica, Universidad de La Laguna, E-38205, Tenerife, Spain\\
$^{3}$Sterrenkundig Observatorium, Ghent University, Krijgslaan 281, S9, B-9000 Ghent, Belgium\\
$^{4}$Laboratoire d'astrophysique, \'Ecole Polytechnique F\'ed\'erale de Lausanne (EPFL), Observatoire, 1290 Versoix, Switzerland\\
}
\date{Published:2016MNRAS.463.3409V}

\pubyear{2016}

\begin{document}
\label{firstpage}
\pagerange{\pageref{firstpage}--\pageref{lastpage}}
\maketitle

\begin{abstract}

We present UV-extended E-MILES stellar population synthesis models covering the
spectral range $\lambda\lambda$ $1680-50000$\,\AA\ at moderately high
resolution. We employ the NGSL space-based stellar library to compute spectra of
single-age, single-metallicity stellar populations in the wavelength range from
$1680$ to $3540$\,\AA. These models represent a significant improvement in
resolution and age/metallicity coverage over previous studies based on earlier
space-based libraries. These model spectra were joined with those we computed in
the visible using MILES, and other empirical libraries for redder wavelengths.
The models span the metallicity range $-1.79\leq\Mh\leq+0.26$ and ages above
$30$\,Myr, for a suite of IMF types with varying slopes. We focus on the
behaviour of colours, spectra and line-strength indices in the UV range as a
function of relevant stellar population parameters. Whereas some indices
strengthen with increasing age and metallicity, as most metallicity indicators
in the visible, other indices peak around $3$\,Gyr for metal-rich stellar
populations, such as Mg at $2800$\,\AA. Our models provide reasonably good fits
to the integrated colours and most line-strengths of the stellar clusters
of the Milky-Way and LMC. Our full-spectrum fits in the UV range for a
representative set of ETGs of varying mass yield age and metallicity
estimates in very good agreement with those obtained in the optical
range. The comparison of UV colours and line-strengths of massive ETGs with
our models reveals the presence of young stellar components, with ages in the
range $0.1-0.5$\,Gyr and mass fractions $0.1-0.5$\%, on the top of an old
stellar population. 

\end{abstract}



%
\section{Introduction}

The study of colours, line-strength indices and spectral energy distributions
(SEDs) of unresolved star clusters and galaxies allows us to determine
fundamental physical properties that are related to their stellar populations,
such as the Star Formation History (SFH), age, metallicity, abundance element
pattern, Initial Mass Function (IMF) and dust properties. Quantifying these
properties is mandatory for an in-depth understanding of the formation and
evolution of these systems. These studies are performed by comparing the
observational data to predictions from stellar population synthesis models
\citep[\eg][]{Tinsley80}. These models are also useful for a wider community,
such as, \eg, for kinematic studies, redshift measurements or constraining
galaxy masses, just to mention some of the possible applications. 

The so called evolutionary population synthesis models, which consider the
relative contributions from all type of stars in the proportions
prescribed by most recent advances in stellar evolution theory, are
nowadays regarded as standard. Most of these models rely on three main
ingredients, i.e. a prescription for the IMF, a set of stellar isochrones
and stellar spectral libraries. The latter, which can either be
theoretical or empirical, allow us to predict a variety of observables in
several bands such as fluxes and colours, mass-to-light ratios (M/L),
surface brightness fluctuations (SBFs), absorption line-strength indices
or spectra with varying resolutions. Such predictive abilities have made
these models increasingly popular
\citep*[\eg][]{Bressan94,Fritze94,Worthey94,Vazdekis96,Pegase,
KodamaArimoto97,Maraston98,Leitherer99,BC03,Schiavon07,ConroyGunn10,Chung13}. 

Most detailed stellar population studies have focused so far on key
absorption line indices, such as those of the Lick system
\citep{Wortheyetal94}. By fitting these indices we are able to partially
lift the main degeneracies affecting the colours, such as that between the
age and the metallicity \citep[\eg][]{Worthey94}, and the abundances of
various elements
\citep*[\eg][]{Rose85,Thomas03,Carretero04,Yamada06,Schiavon07,Conroy14,Worthey14},
without being affected by dust \citep{MacArthur05}. This method has been
mainly applied to early-type galaxies (ETGs) \citep[\eg][]{Trager98}. With
this type of analysis it has been possible to conclude that massive ETGs
have mainly old stellar populations with the bulk of their stars formed at
high redshift in very short timescales \citep[\eg][]{Renzini06}. A new
generation of models predicting spectra of stellar populations at higher
resolutions than is tipically allowed by galaxy dynamics (as a result of
the smearing due to their velocity dispersion), have opened new means to
further improve these studies
\citep[\eg][]{Vazdekis99,Schiavon02,BC03,LeBorgne04,MILESIII,Maraston11,CvD12models}.
This has been possible by employing extensive empirical stellar spectral
libraries, mostly in the optical range
\citep[\eg][]{Jones99,ELODIE01,LeBorgne03,Valdes04,MILESI}, but also
theoretical libraries \citep[\eg][]{Coelho05}. The latter has been
employed for computing spectra of single-age, single-metallicity stellar
populations (SSPs) with varying abundance ratios, either fully theoretical
\citep[\eg][]{Coelho07} or combined with empirical libraries
\citep[\eg][]{Walcher09,Vazdekis15}. These models provide greater
abilities to define and optimize new line indices with enhanced
sensitivities to relevant stellar population parameters
\citep*[\eg][]{Serven05,CervantesVazdekis09}, including IMF indicators
\citep[\eg][]{LaBarbera13,Spiniello14}, and full spectrum-fitting
\citep[\eg][]{Cid05,Ocvirk06a,Koleva09,Tojeiro11}.

Most of these detailed studies have been performed in the optical spectral
range, despite the potential constraining power of the UV and Near-IR, which are
dominated by different types of stars. There are models based on theoretical
stellar libraries that provide us with low resolution SEDs redward the optical
range (e.g., \citealt{BC03,Maraston05,ConroyGunn10}). It has been only recently
that, by means of empirical stellar libraries \citep*[\eg][]{CATI,IRTFI,IRTFII},
we have at disposal SSP spectra at higher resolutions
\citep[\eg][]{CATIV,CvD12models,Meneses15,Roeck15}.  

The lack of libraries extending blueward $\sim3500$\,\AA\ has prevented us to
build models that allow us to perform detailed studies, such as those
performed in the optical range. The UV spectral range is particularly
sensitive to the hot components of galaxy stellar populations
\citep[\eg][]{Faber83}. Therefore, the analysis of the UV places us in better
position to properly characterise galaxy SFHs. Furthermore, as the various
stellar population components contribute in different proportions to the
different parts of the SED the combination of UV and optical (and/or Near-IR)
data allows us to properly separate these contributions
\citep[\eg][]{Rose84,Schiavon04,PercivalSalaris11}. The UV range is full of
absorption features that are dominated by a variety of elements, which further
inform us on the overall metallicity and individual abundance ratios
\citep[\eg][]{Fanelli90,DavidgeClark94,Ponder98,Heap98,Chavez07,Maraston09,Toloba09,Serven11}.
Blue Horizontal Branch (BHB) stars are claimed to be affecting the Far-UV
range of globular clusters \citep[\eg][]{deBoer85,Montes14} or Extreme
Horizontal Branch (EHB) stars as responsible for the UV-upturn phenomenon
observed in a fraction of elliptical galaxies
\citep*[\eg][]{CodeWelch79,Burstein88,Dorman95,Kaviraj07,Yi11,HernandezBruzual14}.
Very importantly, ground-based observing facilities allow us to obtain spectra
of high redshift galaxies whose restframe UV fall in the optical or Near-IR
spectral ranges,  where the current instrumentation is most developed
\citep[\eg][]{Pettini00,deMello04,Cimatti05,Daddi05,Mignoli05,Popesso09,vanDokkumBrammer10}. 

A first attempt to obtain an empirical UV library with a reasonably large range
of stellar parameters was observed by \cite{IUEI} and \cite{IUEII} with the
International Ultraviolet 
Explorer\footnote{\url{http://archive.stsci.edu/iue/}} (IUE). This library is
composed of $218$ stars at resolution $\sim7$\,\AA\ (FWHM), and it has been
employed by the models of \citet{BC03} and \citet{Maraston09}. One of the
shortcomings of the IUE is that the coverage in metallicity is limited to
solar, and so the resulting model predictions.

A significant step further was made by the New Generation Spectral Library
(NGSL) \citep{NGSL}, which was observed with the Imaging Spectrograph (STIS)
onboard the Hubble Space Telescope (HST). The stellar spectra of the NGSL cover
the range $\sim 0.16$--$1.02$\,$\mu$m at resolution R $\sim1000$ and, therefore,
unlike the IUE, it does not reach the Far-UV and it misses Ly\alfa. However the
$374$ stars of this library have a good coverage of stellar atmospheric
parameters, with metallicities between $-2.0$\,dex and $0.5$\,dex and spectral
types from O to M for all luminosity classes. In \citet{NGSLI} we determined the
atmospheric parameters of this library in an homogeneous manner.

Here we make use of the NGSL to extend our stellar population models to the UV
spectral range (Section\,\ref{sec:models}). The computed SSP spectra are joined
to those we predict for the visible (\citealt{MILESIII}, as recently updated in
\citealt{Vazdekis15}) and redder spectral ranges
\citep{MIUSCATI,Roeck15,Roeck16} to cover the range $\lambda\lambda$
$1680$--$50000$\,\AA, at moderately high resolution, all based on empirical
stellar libraries (Section\,\ref{sec:E-MILES}). In Section\,\ref{sec:behaviour}
we focus on the behaviour of the colours, spectra and line-strength indices in
the UV spectral range. In Section\,\ref{sec:comparison} we compare our models
with other predictions in the literature. In Section\,\ref{sec:applications} we
compare our model predictions to stellar cluster and galaxy data. Finally, in
Section\,\ref{sec:summary} we summarize our main results and conclusions.

\section{Model construction}
\label{sec:models}

The models computed in this work employ scaled-solar isochrones and do not
take into account the specific abundance element ratio of the stellar spectra.
Therefore these models, for which we assume that $\Mh=\Feh$, can be considered
"base" models. As the empirical stellar spectra follow the Milky-Way abundance
pattern as a function of metallicity, the resulting base models are nearly
consistent and scaled-solar around solar metallicity. However at low
metallicity they lack consistency as these models combine scaled-solar
isochrones with \alfa-enhanced stellar spectra. Note that in
\citet{Vazdekis15} we computed self-consistent models in the optical range,
which are scaled-solar or \alfa-enhanced for all metallicities, with the aid
of the theoretical stellar spectra of \citet{Coelho05,Coelho07}.

We employ the two sets of scaled-solar theoretical isochrones of \citet{Padova00}
(hereafter Padova00) and \citet{Pietrinferni04} (hereafter BaSTI). The Padova00
isochrones cover a wide range of ages, from $0.063$ to $17.8$\,Gyr, and six
metallicity bins ($Z=0.0004$, $0.001$, $0.004$, $0.008$, $0.019$ and $0.03$),
where $0.019$ represents the solar value. The range of initial stellar masses
extends from $0.15$ to $7$\,\Msol. In comparison to their isochrones published
in \citet{Padova94} the Padova00 employ an improved version of the equation of
state, the opacities of \citet{AlexanderFerguson94} and a milder convective
overshoot scheme. A helium fraction was adopted according to the relation:
$Y\approx0.23+2.25Z$. 

We also use here the BaSTI theoretical isochrones of \citet{Pietrinferni04}
supplemented by additional computations as described in \citet{Vazdekis15},
which include an extra (supersolar) metallicity bin, and extension of the
isochrones to the very low-mass (VLM) regime down to $0.1$\,\Msol, based on the
models of \citet{Cassisi00}. We note that the temperatures for these stars are
cooler than those of Padova00 \citep{MIUSCATI}. For a complete description of the BaSTI database we
refer the interested reader to 
\citet{Pietrinferni04,Pietrinferni06,Pietrinferni09,Pietrinferni13} and
\citet{Cordier07}. We adopted the non-canonical BaSTI models with the mass loss
efficiency of the Reimers law \citep{Reimers} set to $\eta=0.4$. The initial He
mass fraction ranges from $0.245$ to $0.303$, for the more metal-poor to the
more metal-rich composition, respectively, with $\Delta Y / \Delta Z \approx
1.4$. The adopted BaSTI isochrones cover twelve metallicity bins: $Z=0.0001$,
$0.0003$, $0.0006$, $0.001$, $0.002$, $0.004$, $0.008$, $0.0100$, $0.0198$,
$0.0240$, $0.0300$ and $0.0400$. For each metallicity, the isochrone age range
covers the interval from $0.03$ to $14$\,Gyr. We note that for the solar
metallicity model set the atomic diffusion of both helium and metals was
properly accounted for to be able to match accurately the helioseismological
constraints. The best match to the depth of the convective envelope
($0.716$\,\Rsol) of the present solar envelope He abundance (Y$=0.244$), and of
the actual (Z/X) ratio [(Z/X$=0.0244$], lead to an initial He abundance and
metallicity (Y$_{\odot}=0.2734$, Z$_{\odot}=0.0198$). 

There are differences among these two sets of isochrones. The MS loci are in
good agreement. The TO stars have similar luminosity for old stellar
populations, but their luminosities differ for young and intermediate age
regimes. The BaSTI models provide systematically cooler(hotter) RGB at
low(high) metallicity than the Padova00 models. The core He-burning stage is
hotter for the BaSTI models for old populations, mostly due to differences in
the adopted efficiency of mass-loss along the RGB. However, for young ages the
Padova00 models show more extended blue loops, due to differences in the
physical inputs and the treatment of convention of the He-burning core
intermediate-mass stars. Finally, we note that both the Padova00 and BaSTI
isochrones include the thermally pulsing AGB regime using simple synthetic
prescriptions. However the contributions of these red stars are nearly
negligible in the UV spectral range. We refer the interested reader to
\citet{Cassisietal04}, \citet{Pietrinferni04} and \citet{Vazdekis15} for a more
detailed comparison of these two sets of isochrones.

None of these models include stellar rotation, which is particularly
relevant for massive stars. Among other effects rotation brings more material to
the convective core and therefore increases the MS lifetimes by as much as
$\sim25\%$ and decreases the surface gravity and the opacity in the radiative
envelope, raising the luminosity \citep{MaederMeynet00}. As a result of rotation
the colours of the stellar populations with ages smaller than $\sim40$\,Myr vary
by $0.1$ -- $1$\,mag with respect to the non-rotating models, being this
difference larger with increasing wavelength and increasing metallicity
\citep{Vazquez07}. Note that our BaSTI-based models reach $30$\,Myr and
therefore the stellar rotation of massive stars, not included here, might have a
non negligible effect on these young populations. Nevertheless we obtain
reasonably good fits to the absorption line-strengths of LMC stellar clusters in
this age regime, as discussed in Section\,\ref{sec:GCslines}.


The theoretical parameters of these isochrones are transformed to the
observational plane to obtain stellar fluxes using empirical (rather than
theoretical) relations between colours and stellar parameters (\Teff, $\log g$,
\Feh). We use the metallicity-dependent relations of \citet*{Alonso96} and
\citet*{Alonso99} for dwarfs and giants, respectively. Two extensive photometric
stellar libraries of dwarfs and giants (around $\sim500$ stars each library)
were employed to derive these temperature scales throughout the IR-Flux method,
which is only marginally dependent on model atmospheres. We use the empirical
compilation of \citet*{Lejeune97, Lejeune98} (and references therein) for the
coolest dwarfs (\Teff$\lesssim4000$\,K) and giants (\Teff$\lesssim3500$\,K) for
solar metallicity, and also for stars with temperatures above $\sim8000$\,K. We
use a semi-empirical approach for the low temperature stars of other
metallicities. For this purpose we combine these relations and the model
atmosphere predictions of \citet{Bessell89,Bessell91} and the library of
\citet{Fluks94}. We also use the metal-dependent bolometric corrections of
\citet*{Alonso95} and \citet{Alonso99} for dwarfs and giants, respectively, and
adopt $BC_{\odot}=-0.12$. Assuming $V_\odot=26.75$ \citep{Hayes85} we obtain for
the sun the absolute magnitude ${\rm M_{V_\odot}=4.82}$ and ${\rm
M_{{bol}_{\odot}}}$ is given by ${\rm M_{V_{\odot}}+BC_{V_{\odot}}}= 4.70$.

To compute the integrated spectrum of a single-age and single-metallicity
stellar population, SSP, we integrate along the isochrone the stellar spectra
of the NGSL library \citep{NGSLI}, which is described in
Section\,\ref{sec:NGSL}. Specifically, we follow the approach described in
\citet{Vazdekis15} for the base models. In this integration the number of
stars in each mass bin comes from the adopted IMF, for which we assume four
shapes as summarized in \citet{CATIV,Vazdekis15}. These include the multi-part
power-law IMFs of \citet{Kroupa01}, i.e. universal and revised, and the two
power-law IMFs described in \citet{Vazdekis96}, i.e unimodal and bimodal, both
characterised by the logarithmic slope, $\Gamma$ and $\Gamma_b$, respectively,
as a free parameter. In addition, in this work we also have implemented the
\citet{Chabrier} single-stars IMF. The \citet{Salpeter55} IMF is obtained by 
adopting the unimodal IMF with $\Gamma=1.35$, and (although not identical) the 
Kroupa Universal IMF is very similar to a bimodal IMF with slope $\Gamma_b=1.3$. 

During the integration we scale the stellar spectra according to their fluxes
in the $V$ broad-band filter, which is fully contained in the NGSL stellar
spectra. The predicted flux in this band is computed according to the same
empirical photometric relations employed for the isochrones. To normalise the
spectra we convolve with the filter response of \citet{BuserKurucz78}.  To
obtain the absolute flux in the $V$-band we follow the method described in
\citet{FalconBarroso11}, which is based on the calibration of
\citet*{Fukugita95}. We obtain the zero-point using the Vega spectrum of
\citet{Hayes85} with a flux of $3.44\times10^{-9}$\,erg
cm$^{-2}$\,s$^{-1}$\,\AA$^{-1}$ at 5556\,\AA, and the $V$ magnitude set to
$0.03$\,mag, consistent with \citet{Alonso95}.

For each star, with a given set of atmospheric parameters, we use the spectra
of the stars of the library with the closest parameters to apply the
interpolation scheme described in \citet{CATIV} (see Appendix B), as updated
in \citet{Vazdekis15}. This algorithm selects the stars whose parameters are
within a box around the requested parametric point (\Teff, \logg, \Feh),
which is divided in eight cubes, all with one corner at that point. If no
stars are found in any of these cubes it can be expanded until suitable stars
are found. This local interpolation algorithm is particularly suitable to
overcome the limitations inherent to the gaps and asymmetries in the
distribution of stars around that point. The larger the density of stars
around it the smaller the box is, which can be as small as imposed by typical
uncertainties in the determination of the parameters (see \citealt{CATII}).
We assign the smallest box to those parametric points for which the NGSL
provided density values where the cumulative density fraction reached the
upper $99.7$ percentile. For example, this has been the case for stars with
similar parameters to the sun, or solar metallicity giants with temperatures
around $4800$\,K. In each of these cubes, the stars are combined taking into
account their parameters and the signal-to-noise of their spectra. Finally,
these combined stars are weighted to obtain the stellar spectrum with the
requested atmospheric parameters. 

It is worth noting that, within this integration scheme, a stellar spectrum
is attached to the isochrones following the requested stellar parameters (\Teff,
\logg, \Feh), irrespective of the evolutionary stage. This is particularly
relevant for the AGB phase in the intermediate-age regime, although these stars
are more influential at redder wavelength ranges. More relevant to the UV is the
evolution of binary (or multiple) stars, which we do not take into account in
the models presented here. Furthermore we have decreased, and in some cases
discarded, the spectroscopic binaries that are present in the library (see
Section\,\ref{sec:NGSL}). However we provide models computed with the Kroupa
Revised IMF \citep{Kroupa01}, which correct the relative number of stars for the
systematic bias due to unresolved binary proportion. Very recently
\citet{HernandezBruzual14} implemented in their stellar population models
varying fractions of interacting binary pairs as progenitors of EHB stars to
reproduce both UV-weak and UV-strong ETGs. In contrast, e.g., \citet*{Smith12},
favour the metal-rich single-star origin to explain the UV flux excess in old
stellar populations. The correlation found between the Far-UV colour and
metallicity comes naturally within this scenario, as mass loss during the RGB
phase increases with metallicity producing blue HB stars. To be able to properly
contribute to this discussion we would need to extend the models to the Far-UV,
which is mostly out of the reach of the spectral range covered by the NGSL
stellar spectra implemented here. Note however that the IUE stellar spectral
library is very useful for this purpose.

\subsection{The stellar spectral library}
\label{sec:NGSL}

The main ingredient employed for extending our models to the UV spectral
range is the NGSL\footnote{\url{http://archive.stsci.edu/prepds/stisngsl/}}
\citep{NGSL}. The NGSL stars were observed with STIS on-board HST using
three gratings (G230LB, G430L and G750L). The obtained spectra for each star
overlap at $2990-3060$\,\AA\ and $5500-5650$\,\AA. The final spectra cover
the wavelength range from $\sim0.16$ to $\sim$1.02\,$\mu$m (slightly different
from star to star). The flux-calibration reaches a precision of $3$\,percent
\citep{Heap09}.  The targeted stars ($600$) were chosen to sample four
metallicity groups, with roughly 150 stars in each bin: \Feh$ < -1.5$; $ -1.5 <
\Feh < -0.5$; $ -0.3 < \Feh < +0.1$; $ +0.2 < \Feh$. However, about $200$
stars were not observed due to the STIS failure in $2004$.

In \citet{NGSLI} we downloaded version $2$ of the reduced data to determine in
an homogeneous way the atmospheric parameters of the NGSL stars. For
determining the stellar parameters we used the full spectrum fitting package
\ulyss\ \citep{Koleva09}. We reached a precision for the FGK stars of $42$\,K,
$0.24$ and $0.09$ dex for \Teff, \logg\ and \Feh, respectively. The obtained
mean errors for the M stars were $29$\,K, $0.50$ and $0.48$ dex, and for the
OBA stars $4.5$\,percent, $0.44$ and $0.18$\,dex. In \citet{NGSLI} we also
determined the  wavelength calibration precision and characterised the
intrinsic resolution of the NGSL spectra.  We find that the wavelength
calibration is precise up to $0.1$\,px, after correcting a systematic effect
in the optical range.  The spectral resolution varies from $3$\,\AA\ in the UV
to $10$\,\AA\ in the near-infrared (NIR), corresponding to a roughly constant
reciprocal resolution $R=\lambda/\delta\lambda \approx1000$ and an
instrumental velocity dispersion $\sigma_{ins} \approx 130$\,\kms. 

In \citet{NGSLI} we estimated the Galactic extinction for each of the
stellar spectra. In brief, this was perfomed when deriving the fundamental
parameters of the NGSL stars as we fit their spectra with interpolated templates
(e.g. MILES stars) together with a multiplicative polynomial. As both the flux
calibration of the employed spectra and the derived atmospheric parameters are
accurate, we use this polynomial to estimate the Galactic extinction, which is
obtained by fitting the multiplicative polynomial to the \citet{Fitzpatrick99}
law.  To correct from Galactic extinction the individual stellar spectra we
divided them by this fit, and extended it to the blue and red wavelengths. We
refer the interested reader to \citet{NGSLI} for an extensive description of the
method. 	

To be implemented in our models we rebinned the original spectra, with 
varying sampling, namely 
$1.373$\,\AA/px ($\lambda\lambda\ 1675 - 3060\,\AA$),
$2.744$\,\AA/px ($\lambda\lambda\ 3060 - 5650\,\AA$) and
$4.878$\,\AA/px ($\lambda\lambda\ 5650 - 10196\,\AA$) to
a constant value of $0.9$\,\AA/px, i.e., similar to MILES, for the whole spectral 
range. The spectra are calibrated in air wavelengths. Note however that
only the first two spectral ranges are considered here for the models
described in Section\,\ref{sec:E-MILES}.

   \begin{figure}
    \includegraphics[width=0.49\textwidth]{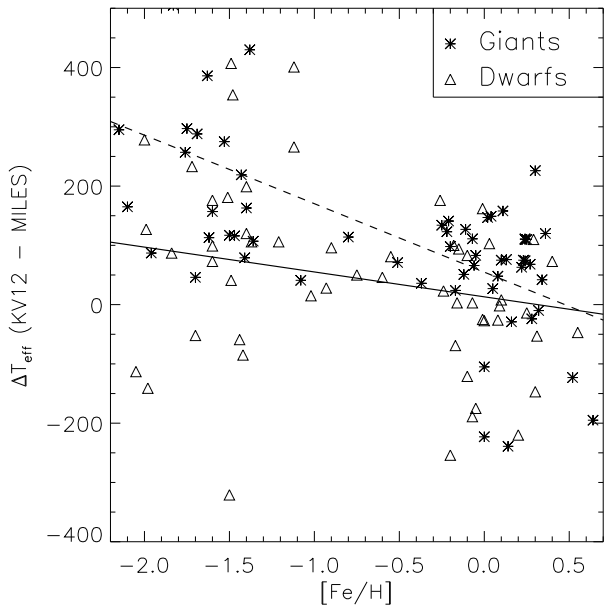}
    \caption{Temperature difference obtained by comparing the values of
    \citet{NGSLI}, quoted as KV12, with those of MILES \citep{MILESII} as a
    function of metallicity for the stars in common between these two libraries.
    The obtained fits for dwarf (open triangles) and giant (asteriscs) stars are
    shown in solid and dashed lines, respectively (see
    Eq.\,\ref{Eq:transformations}).}
    \label{fig:miles_transform}%
   \end{figure}
 
All the stars were visually inspected, one by one, by comparing them to stars
with similar parameters and to the output of the interpolator, once excluding
the target spectrum, following the method described in \citet{MILESIII}. The two
approaches allowed us to identify stars whose spectra clearly deviate from the
expected shape, either globally or within some specific wavelength range. We
employed for the models $303$ stars from the original sample composed of $374$
stars. Nearly half of the removed stars are spectroscopic binaries that were
found to be redundant, as there are other stars with very similar parameters in
the library. Other reasons for removing stars include strong emission along the
whole spectral range, unexpected dips in flux, wrong spectrum shape for the
estimated parameters and some very peculiar stars.  Finally, instead of removing
$15$ stars, whose spectra were partially affected by some of these problems, we
decreased their weight for stellar population modeling by means of the
interpolation scheme we applied to synthesize the requested stars. Among
them are hot stars showing interstellar absorption lines affecting, e.g., the
Mg{\sc II} feature at $\sim2800$\,\AA, as the modest resolution of the NGSL
spectra does not allow us to properly deal with them. This has also been the
case for very cool stars showing chromospheric emission filling in this feature
(e.g. \citealt{Genova90}). This choice of decreasing the weight of these stars
is mainly motivated by the fact that their parameters were difficult to match by
the remaining, cleaner, stars, and we wanted to avoid increasing the number of
local gaps in our stellar parametric distribution. 

We also performed cosmetic corrections to the spectra of $53$, mostly red, 
stars. In many cases where we found some negative flux values, typically
blueward $\sim2150$\,\AA\ where the S/N gets very low (in all cases $<20$ 
for the spectrum obtained with the blue grating of STIS, i.e.
$\lambda<3060$\,\AA), we assigned the minimum positive flux value within a
$20$\,\AA\ wide window that is centered on that pixel(s). For $20$ of these
stars, i.e. mostly the coolest ones, we found that significant portions of
their spectra blueward $\sim2150$\,\AA\ showed too noisy spectrum, including
many negative flux values. Therefore we just assigned the mean flux measured
in the neighbouring, redder, $100$\,\AA\ wide window, which is located
between $2150$ and $2400$\,\AA, depending on each star. Note that the
contributions of these stars in the UV spectral region is negligible.

To match our NGSL-based models to those computed in the optical range (see
Section\,\ref{sec:E-MILES}), we need to homogenize the \citet{NGSLI} stellar parameters
with our reference values adopted in the MILES library \citep{MILESII}. For this
purpose we compared the parameters of the NGSL stars in common with MILES.
Fig.\,\ref{fig:miles_transform} shows the difference in temperature obtained as
a function of metallicity for the dwarf and giants stars. Although there is a
significant scatter throughout the whole metallicity range, we see that the
temperature difference increases with decreasing metallicity and that this
effect is larger for the giants (we refer the interested reader to \citealt{NGSLI} for
further details on these differences). The obtained fits
allowed us to transform these temperatures to the MILES system by applying
the following corrections:

\begin{eqnarray}
{\rm T_{MILES}} &=& {\rm T_{KV12}}-13.1784-41.9432 \Feh ~~~~(\rm dwarfs) \nonumber \\
{\rm T_{MILES}} &=& {\rm T_{KV12}}-54.7807-115.537 \Feh ~~~~(\rm giants)
\label{Eq:transformations}
\end{eqnarray}

For $109$ stars in common with MILES and $15$ stars in common with the CAT
library \citep{CATII}, which is also in the same system as MILES, we adopted the
parameters provided by these libraries. For $168$ stars we corrected the
parameters of \citet{NGSLI} according to Eq.\,\ref{Eq:transformations}.  Finally, for
$10$ stars in common with MILES (HD\,10380, HD\,37828, HD\,44007, HD\,95735,
HD\,112413, HD\,123657, HD\,164353, HD\,167006, HD\,175865, BD\,+37\,1458 and
BD\,+44\,2051) and one in common with CAT (BD\,+44\,205) we did not use the
parameters listed in these libraries as we found that those of \citet{NGSLI}, transformed
in this way, matched better other stars with similar parameters.

  \begin{figure}
   \includegraphics[width=0.49\textwidth]{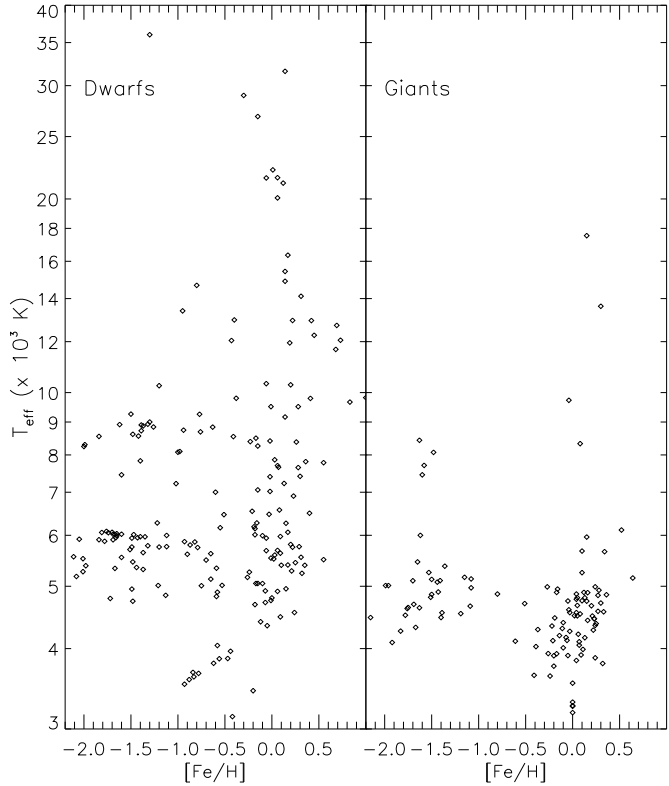}
   \caption{Adopted atmospheric parameters for the NGSL stars. Dwarf stars are
   shown in the left panel whereas the giants are shown in the right panel.}
   \label{fig:atm_params}%
  \end{figure}

The resulting stellar parameters are shown in Fig.\,\ref{fig:atm_params}.
Dwarf stars with temperature in the range $4500-9000$\,K are well covered
within the metallicity range $-1.7<\Feh<+0.3$. The poor coverage of cool
dwarfs is not a problem, since their contribution to the total light in the
UV is small. Similarly the number of hot dwarfs decreases significantly with
increasing temperature, limiting the reliability of our model predictions for
ages smaller than $\sim0.5$\,Gyr. We also see that the highest density of
stars is obtained for solar metallicity. This also applies to the giants as
illustrated by the right panel of the figure.  Note that for some of the
coolest giant stars with metallicity estimates around $-1$\,dex we do not
take their metallicity into acccount. Instead we set their metallicity to 0.0
to optimize the interpolation in this sparsely populated very low temperature
regime. However there is a large gap in the coverage of giants with
\Feh$<-0.3$, with some improvement around \Feh$<-1.7$. This gap prevents us
of building stellar population models in spectral regions dominated by these
redder stars. The model limitations imposed by these gaps are discussed in
detail in the next section.

\subsection{Quality of the UV SSP spectra}
\label{sec:quality}

To estimate the quality of the SSP spectra in the UV range we follow the
quantitative approach introduced in \citet{MILESIII}, which makes use of the
interpolation algorithm described in Section\,\ref{sec:models}. We compute a
normalised quality parameter, $Q_n$, which is related to the density of stars
around the atmospheric parameters of the stars that are requested when
integrating along the isochrone. Basically, the higher these densities are in
the NGSL, the higher the $Q_n$ value of the resulting SSP spectrum. To obtain
$Q_n$ we normalise it with respect to a minimum acceptable value, which comes
from a poor, but still acceptable, parameter coverage. We refer the interested
reader to \citet{MILESIII} for a full description of the method. The main
difference, apart of the use of the NGSL, is that we use the flux in the U band
for the weighting of the stars required to compute this parameter (see Equation
5 of \citealt{MILESIII}). This is motivated by the fact that we only will use
the UV spectral region of these models (see Section\,\ref{sec:E-MILES}).

\begin{figure}
   \centering
  \includegraphics[width=0.49\textwidth]{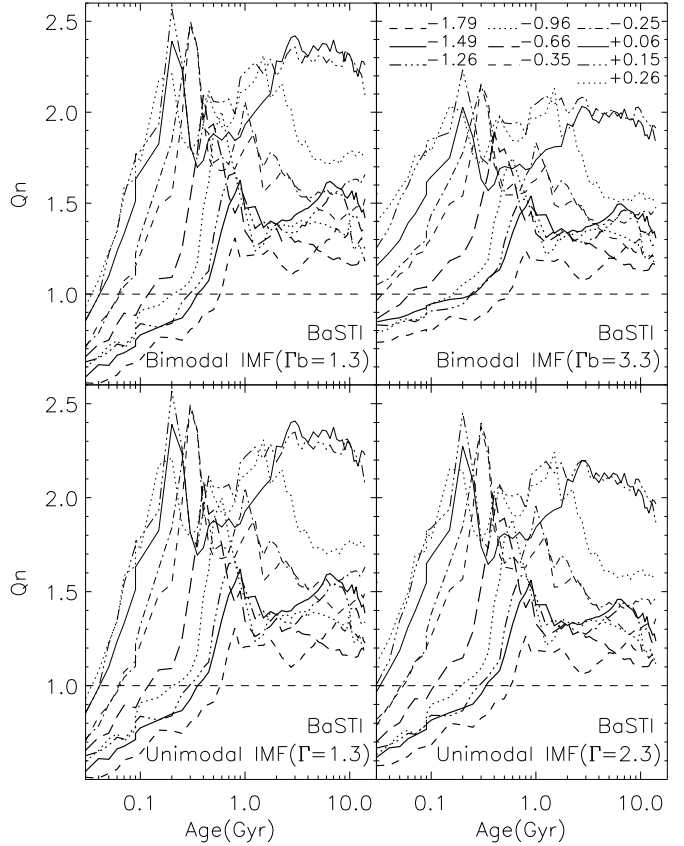}
   \caption{The normalized quality parameter, Q$_n$, of the NGSL based SSP spectra
   computed with BASTI isochrones is plotted as a function of age for various
   metallicities as indicated within the upper-right panel. Models can be
   considered safe for Q$_{n}>1$. The upper-left panel shows the quality of the
   models computed with a bimodal IMF with slope 1.3, whereas a rather
   bottom-heavy IMF with slope 3.3 is shown in the upper-right panel. The Q$_{n}$
   values corresponding to the models computed with a Unimodal IMF with slope
   1.3 (i.e. Salpeter) and 2.3 are shown in the bottom-left and bottom-right
   panels, respectively.}
   \label{fig:Qngsl_BASTI}%
\end{figure}

\begin{figure}
   \centering
  \includegraphics[width=0.49\textwidth]{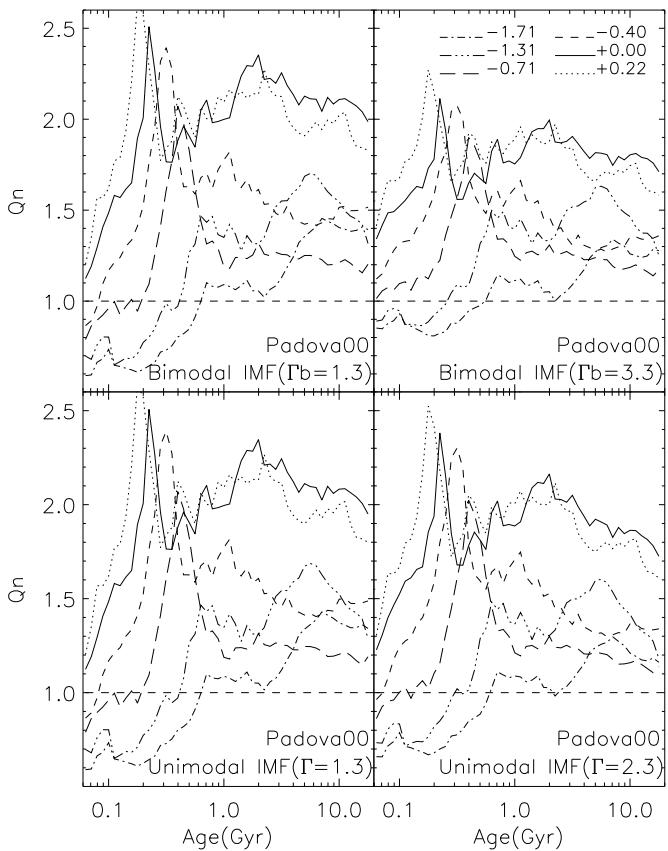}
   \caption{Same as Fig.\,\ref{fig:Qngsl_BASTI}, but for the models computed with
   the Padova00 isochrones.}
   \label{fig:Qngsl_Padova00}%
\end{figure}

Fig.\,\ref{fig:Qngsl_BASTI} and Fig.\,\ref{fig:Qngsl_Padova00} show the
resulting $Q_n$ values for the models computed on the basis of the BaSTI and
Padova00 isochrones. Models for which $Q_{n}>1$ can be considered acceptable,
whereas the ones with values slightly lower than one can still be used with
some caveats, depending on the applications such as, e.g. colours. These two
figures show that the models are safe for all the plotted metallicities for
ages above $\sim0.5$\,Gyr. For metallicities around solar these models are
safe for younger ages, particularly when steepening the IMF, as illustrated in
the right panels. For intermediate-aged stellar populations we observe a
relative increase of the $Q_{n}$ values as a result of the fact that in our
interpolation scheme we do not consider the metallicity of the stars for
temperatures above $\sim9000$\,K. The largest $Q_{n}$ values are obtained for
the models with metallicities around solar. The quality of the models with
\Mh$\sim0.2$ drops significantly due to the poorer coverage of stars with
supersolar metallicities in the NGSL. The quality of the models also decreases
for low metallicities. It is not surprising that for the old stellar
populations the lowest $Q_{n}$ values are obtained for metallicities around
\Mh$\sim-0.7$. This result is consistent with the scarce coverage of stars
with metallicities around such value, as shown in Fig.\,\ref{fig:atm_params}.

Fig.\,\ref{fig:Qngsl_BASTI} and Fig.\,\ref{fig:Qngsl_Padova00} also show a
peculiar behaviour of the $Q_{n}$ parameter as a function of IMF slope. For
intermediate- and old-aged stellar populations the quality decreases slightly
when steepening the IMF. In contrast, for the  models with the youngest ages
a bottom-heavier IMF contributes to increase the value of the quality
parameter. Such a behaviour can be understood by tracking the density of
stars along the temperature axis in the left panel of
Fig.\,\ref{fig:atm_params}. For old stellar populations and bottom-heavy IMFs
the contribution to the total light of MS dwarfs with \Teff$\lesssim4500$\,K
increases relatively to the hotter stars, compared to a standard IMF. As the
density of such cool stars is lower in the NGSL library, the resulting
$Q_{n}$ values decrease. On the contrary, for the stellar populations with
the youngest ages the relative contribution of stars with
\Teff$\gtrsim10000$\,K (less abundant) with respect to the MS stars of lower
temperature (more abundant) decreases when the IMF becomes bottom-heavier. 

\begin{figure}
   \centering
  \includegraphics[width=0.49\textwidth]{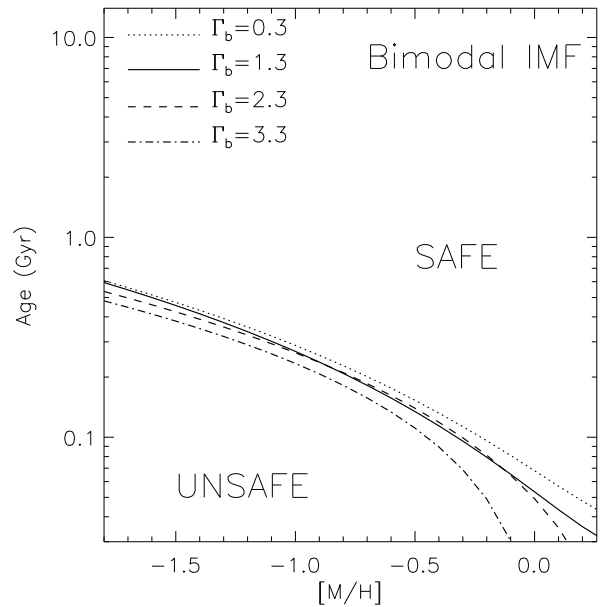}
   \caption{Safe and unsafe models with bimodal IMF and varying slopes as marked
   within the plot. The safe age and metallicity ranges showed here refer to the
   UV spectral range and apply to the models computed with the two sets of
   isochrones feeding our models. See the text for details.}
   \label{fig:QU_unsafe}%
\end{figure}

\begin{figure}
   \centering
  \includegraphics[width=0.49\textwidth]{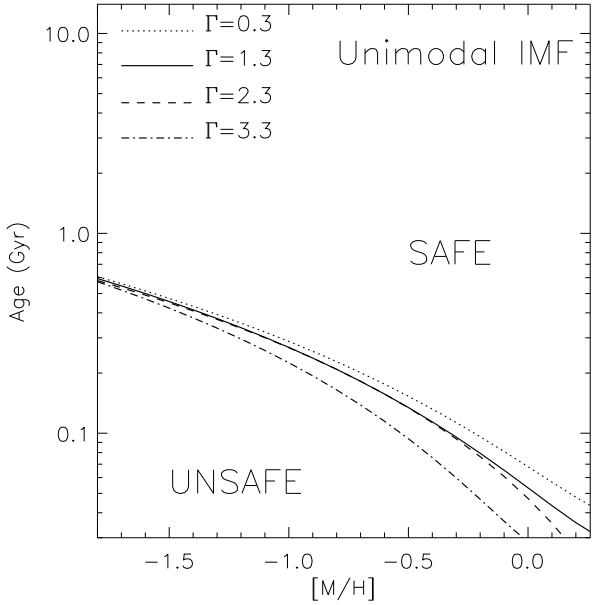}
   \caption{Same as Fig.\,\ref{fig:QU_unsafe}, but for models adopting a
   unimodal IMF with various slopes as quoted within the plot.}
   \label{fig:QUun_unsafe}%
\end{figure}

Finally, in Fig.\,\ref{fig:QU_unsafe} and Fig.\,\ref{fig:QUun_unsafe} we
summarize the safe and unsafe SSPs in the UV spectral range for the models with
bimodal and unimodal IMF types, respectively. The plotted curves, which
correspond to different IMF slopes, were obtained by fitting the
age/metallicity corresponding to models with Q$_n$ values around $1$ and
therefore these ranges must be taken with some caveats. Note that the Salpeter
IMF safe/unsafe ranges are well represented by the curve corresponding to the
unimodal case with slope $1.3$, whereas the ones for the Kroupa and Chabrier
IMFs are very close to the bimodal case with slope $1.3$.

\section{E-MILES: the Extended MILES models}
 \label{sec:E-MILES}%

In Section\,\ref{sec:models} we have described how we compute the models in the
NUV making use of the NGSL stellar library, which extend to the red out to
$\sim1\mu$m. However we only use from these models the spectral region blueward
3540\,\AA, which is joined to the blue end of the SSP spectra based on the MILES
stellar library. These models have been recently extended to the near-IR
\citep{Roeck16} by joining the MIUSCAT SSP spectra \citep{MIUSCATI}, which apart
of MILES also employ the Indo-US \citep{Valdes04} and CaT \citep{CATI} stellar
libraries, to the SSP spectra computed on the basis of the IRTF stellar library
\citep{IRTFI,IRTFII} reaching out to 5\,$\mu$m \citep{Roeck15}. All these models
were computed with our population synthesis code in a fully consistent manner.
The chief aspect of these models is that the
resulting SSP spectra are based on empirical stellar spectral libraries
throughout the whole spectral range, $\lambda\lambda$ 1680.2\,\AA--49999.4\,\AA,
at moderately high resolution.

\begin{figure}
   \centering
  \includegraphics[width=0.49\textwidth]{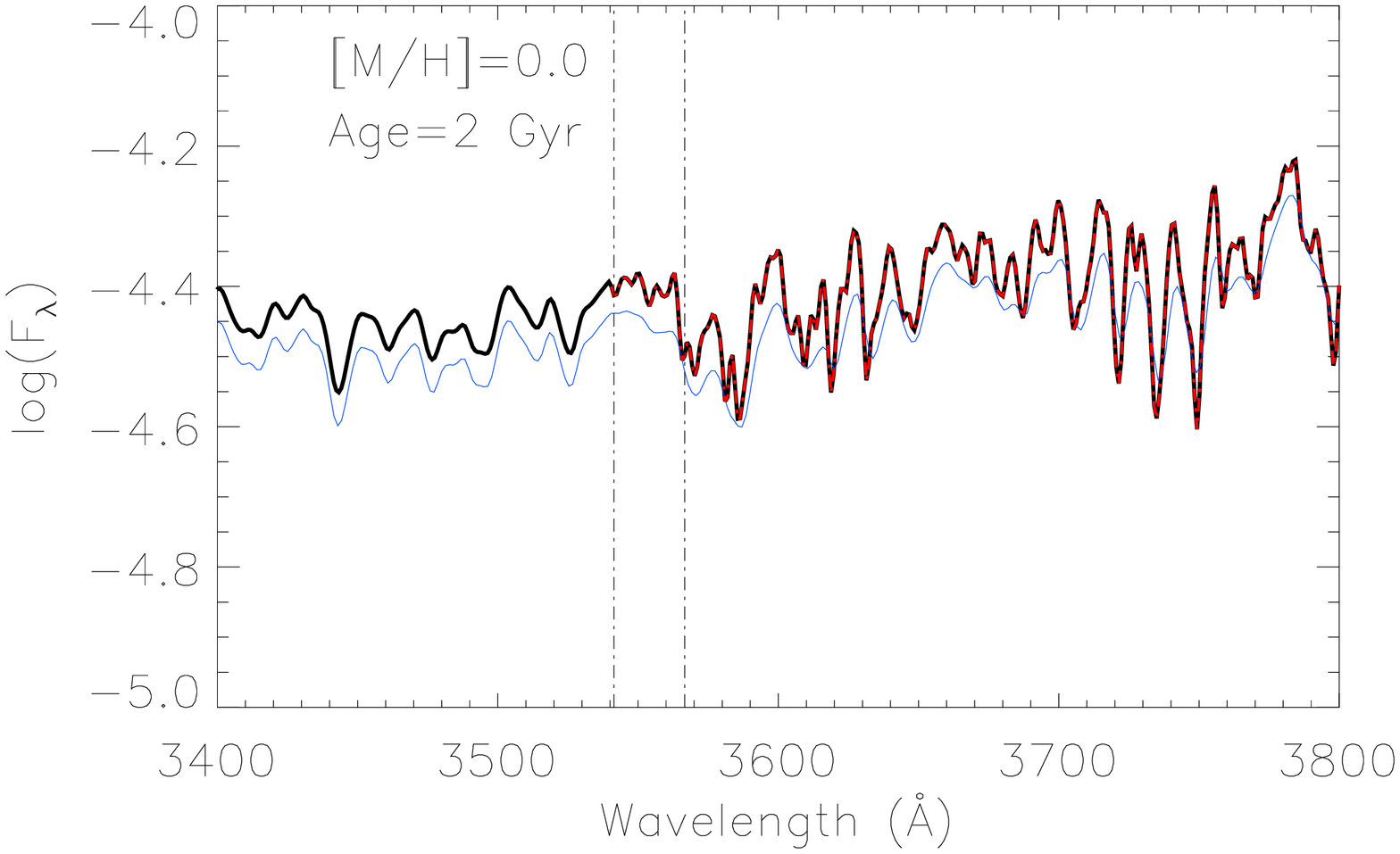}
  \includegraphics[width=0.49\textwidth]{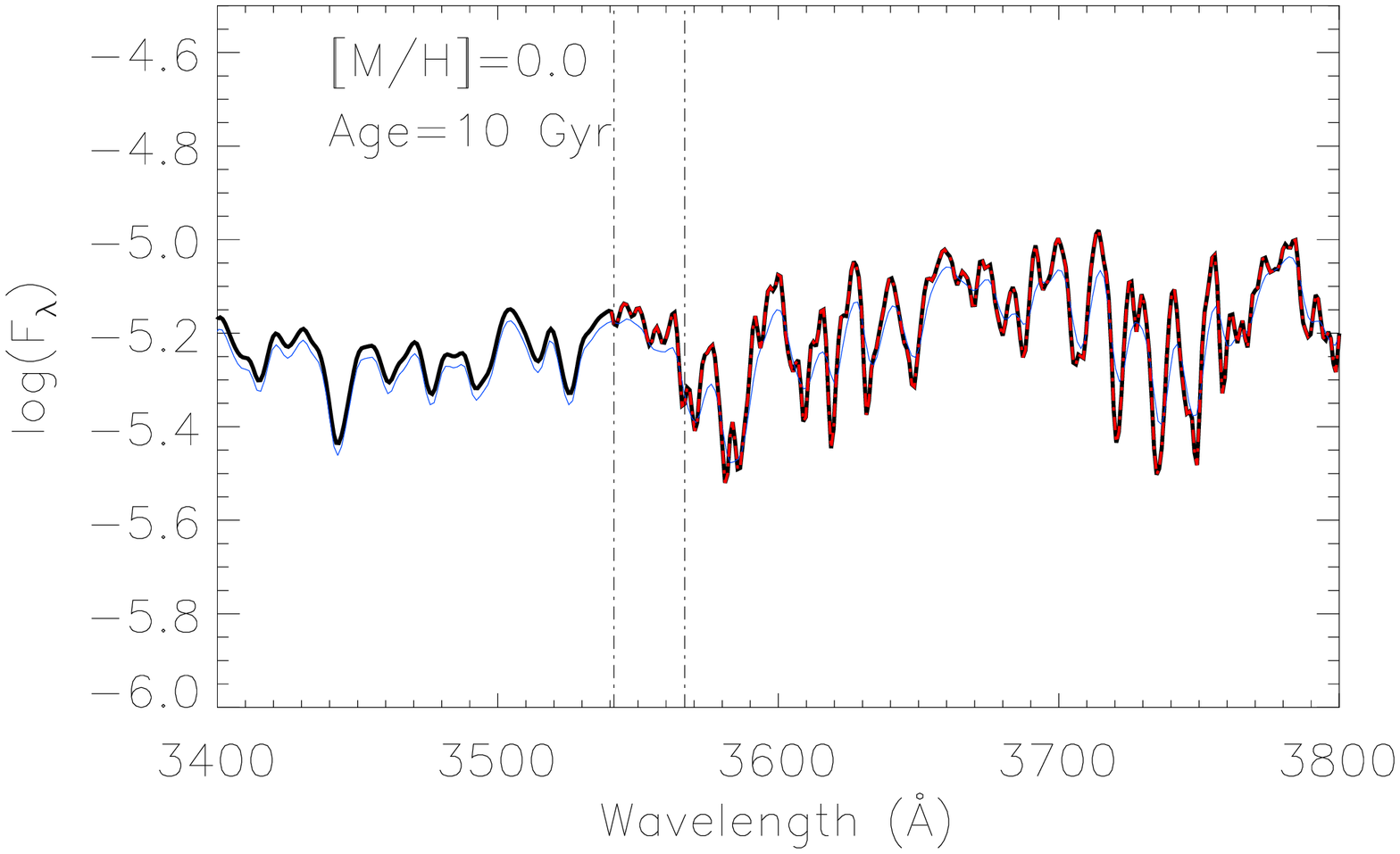}
   \caption{The vertical lines mark the selected pseudo-continuum for
   joining the NGSL- and MILES-based models. The thick black line represents
   the combined SSP spectrum and the thin blue (thick red) ones represent the NGSL (MILES)
   SSP. The panels show two representative SSP ages, 2 (upper panel) and 
   10\,Gyr (lower panel) and solar metallicity. }
   \label{fig:bridge}%
\end{figure}

\begin{table*}
\centering
\caption{\label{tab:SEDproperties}Spectral properties of the E-MILES models.}
\begin{tabular}{ll}
\hline                   
Units &L$_{\lambda}$/L$_{\odot}$\AA$^{-1}$M$_{\odot}$$^{-1}$, L$_{\odot}=3.826\times10^{33}{\rm erg.s}^{-1}$ \\
Continuum shape    & Flux-scaled \\
Spectral range     & $\lambda\lambda$ $1680.2-49999.4$\,\AA\ (air-wavelengths)\\	
Linear dispersion  & $0.9$\,\AA/pix\\
Spectral resolution& FWHM$=3.0$\AA~($1680.2-3060.8$\AA\,)$^{a}$\\    
                   & FWHM$=5.0$\AA~($3060.8-3541.4$\AA\,)$^{a}$\\    
                   & FWHM$=2.51$\AA~($3541.4-8950.4$\AA\,)$^{b,c,d,e}$\\	
                   & $\sigma=60$\kms~($8950.4-49999.4$\AA\,)$^{f,g}$\\	
\hline
$^{a}$This work\\
$^{b,c,d,e}$\citet{CATIV,MILESIII,MIUSCATI,Vazdekis15}\\
$^{f,g}$\citet{Roeck15,Roeck16}\\
\end{tabular}
\end{table*}              

\begin{table*}
\centering
\caption{\label{tab:windows}Spectral windows employed for joining the various
ranges of the E-MILES models}
\begin{tabular}{llll}
Window (\AA) & Spectral resolution variation & Modified spectrum shape & Reference\\
\hline                   
$3541.4-3566.6$ & From FWHM$=5$\,\AA\ to FWHM$=2.5$\,\AA\     & No, only flux
shift applied & This work	 \\
$7360.0-7385.0$ & FWHM$=2.5$\,\AA\ constant   & Yes & \citet{MIUSCATI}\\
$8390.0-8415.0$ & FWHM$=2.5$\,\AA\ constant   & Yes & \citet{MIUSCATI}\\
$8950.0-9100.0$ & From FWHM$=2.5$\,\AA\ to $\sigma=60$\,\kms\ & Yes & \citet{Roeck16} \\
\end{tabular}
\end{table*}              

To join our newly synthesized NUV SSP spectra to those computed in the MILES
range we have followed a similar procedure to the one adopted for building the
MIUSCAT SSP spectra \citep{MIUSCATI}. For this purpose we identified a
spectral region $\lambda\lambda 3541.4$--$3566.6$\,\AA, where no major features
are found for the range of ages and metallicities covered by our
models\footnote{For the E-MILES models we avoided to use the small spectral
range extension blueward MILES, down to $3464.9$\,\AA, of the MIUSCAT models
based on the Indo-US stellar  library.}. This overlapping window has been
chosen to be sufficiently wide to reach enough statistics for the continuum
counts and, at the same time, to avoid the presence of strong spectral
features. To join the two models, we have re-scaled the NGSL based SSP spectra
to match the continuum of the models based on MILES within this
pseudo-continuum window. To compare the two fluxes in this window, we smoothed
the MILES SSP spectra to 5\,\AA\ to match the nominal resolution of our NGSL
based models in this specific spectral range (see \citealt{NGSLI}). Note that
in the combined models we do not change the nominal resolutions of the NGSL
and MILES based SSP spectra.  A zoom of the combined models in the matching
region is shown in Fig.\,\ref{fig:bridge} for two representative SSPs. Note
that the flux calibration of NGSL and MILES models are in good agreement
(within 0.02\,mag), particularly for the old SSP model (lower panel).
Therefore the joining is possible by only slightly adjusting fluxes in the
selected overlapping window. This has been possible due to the good flux
calibration of the stellar spectra of both MILES and the NGSL in this spectral
range. Thus, we ended up covering the wavelength range from $1680.2$ to
$49999.4$\,\AA. Note the drop in resolution from $5$ to $2.51$\,\AA, blueward
$3541.4$\,\AA, is clearly visible in Fig.\,\ref{fig:bridge}.

\begin{figure}
   \centering
  \includegraphics[width=0.49\textwidth]{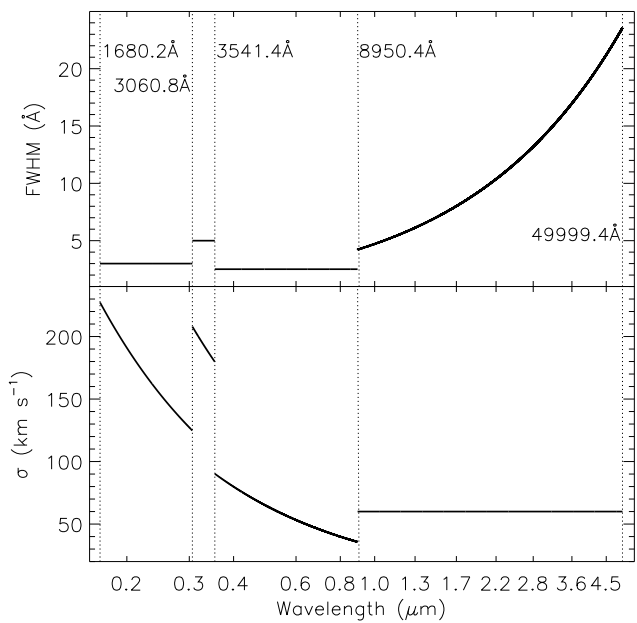}
   \caption{E-MILES spectral resolution as a function of wavelength both in FWHM
(upper panel) and $\sigma$ (lower panel). The vertical dotted lines indicate the
exact wavelength values where the resolution changes. Versions of these models 
both at their nominal resolutions and smoothed to match a constant resolution 
(FWHM and $\sigma$) are available on the MILES website.}
   \label{fig:emiles_resolution}%
\end{figure}

The E-MILES models are computed for both BaSTI and Padova00 isochrones and five 
IMF shapes: Kroupa Universal, Revised Kroupa, Chabrier, Unimodal and Bimodal. For
the latter two functional forms the slope has been varied from very top-heavy
IMF slope ($0.3$) to very bottom-heavy ($3.3$). Table\,\ref{tab:SEDproperties}
lists the resolutions of the various spectral ranges covered by the combined
models. Note that blueward $8950$\,\AA\ the models have a constant FWHM,
whereas for redder wavelengths $\sigma$ is constant.
Fig.\,\ref{fig:emiles_resolution} illustrates the resolution of the E-MILES
spectra as a function of wavelength, both in FWHM and $\sigma$. All these models
can be retrieved from the MILES website \url{http://miles.iac.es}, which also
includes versions of these models smoothed to match a constant resolution (FWHM 
and $\sigma$) along the covered spectral range. 

The quality parameter of the models in the spectral region blueward the MILES
range has been discussed in Section\,\ref{sec:quality}. To assess the
reliability of the models in the various spectral ranges we refer the reader
to the specific papers listed in Table\,\ref{tab:SEDproperties}. The E-MILES
SSP spectra can be considered safe along the whole spectral range as
restricted by the quality of the models redward $8950$\,\AA. These models
employ $180$ carefully selected and prepared empirical stellar spectra from
the IRTF library, which encompass cool supergiants and dwarfs as well as AGB
and some carbon stars. However since the stars of the IRTF library are mostly
bright, nearby stars of around solar metallicity, these models are most
accurate in the metallicity range $-0.40<$\Feh$<+0.20$ for ages above
$1.5$\,Gyr. 

There is a very important caveat to take into account for those users of E-MILES
SEDs who aim at measuring or defining line-strength indices. It should be
avoided line indices whose feature/pseudocontinuum bandpasses are located at
either side of the joining spectral regions corresponding to the various SSP
spectra employed for the combined E-MILES models. The index strength could vary
as a result of the flux correction applied to match the two spectral regions. In
addition it has to be taken into account that for some of these cases the
resolution changes. These joining regions are located around $\sim3540, 7410,
8350$ and $8950$\,\AA. Table\,\ref{tab:windows} lists these spectral windows
with the corresponding resolutions, as well as an indication on whether the
spectrum shape has been modified.

\begin{figure*}
   \centering
  \includegraphics[width=0.9\textwidth]{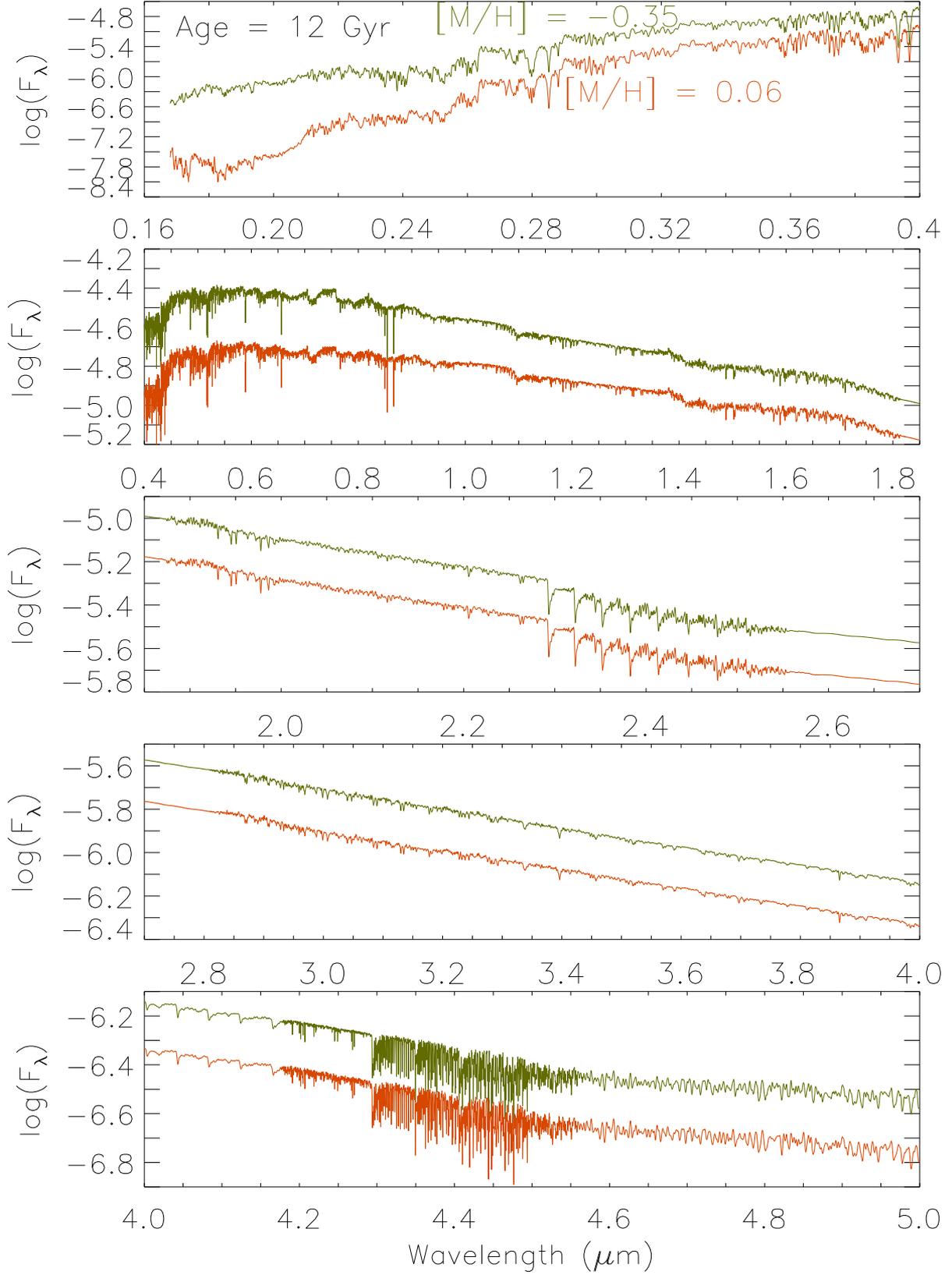}
   \caption{E-MILES SSP spectra with age $12$\,Gyr and Kroupa Universal IMF,
   computed with BaSTI, for two metallicity values: \Mh$=+0.06$ (orange) and
   \Mh$=-0.35$ (olive green). The model with subsolar metallicity has been
   shifted by $+\Delta\log{F_{\lambda}}=1.5$ for visibility.}
   \label{fig:emiles_12Gyr}%
\end{figure*}

\begin{figure*}
   \centering
  \includegraphics[width=0.9\textwidth]{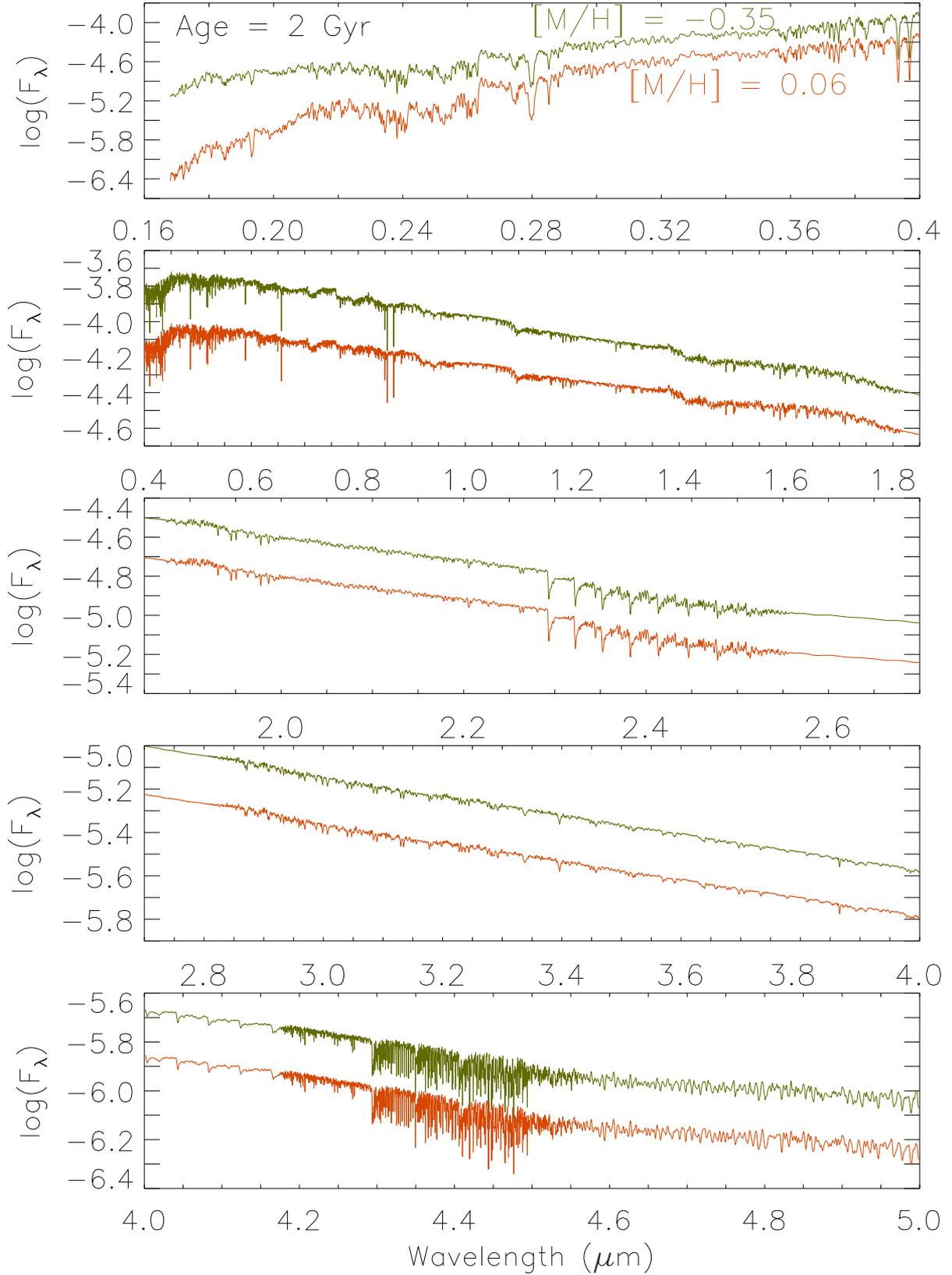}
   \caption{Same as in Fig.\,\ref{fig:emiles_12Gyr} but for models with age
   $2$\,Gyr. The model with subsolar metallicity has been shifted by
   $+\Delta\log{F_{\lambda}}=1.5$ for visibility.}
   \label{fig:emiles_2Gyr}%
\end{figure*}

\begin{figure*}
   \centering
  \includegraphics[width=0.9\textwidth]{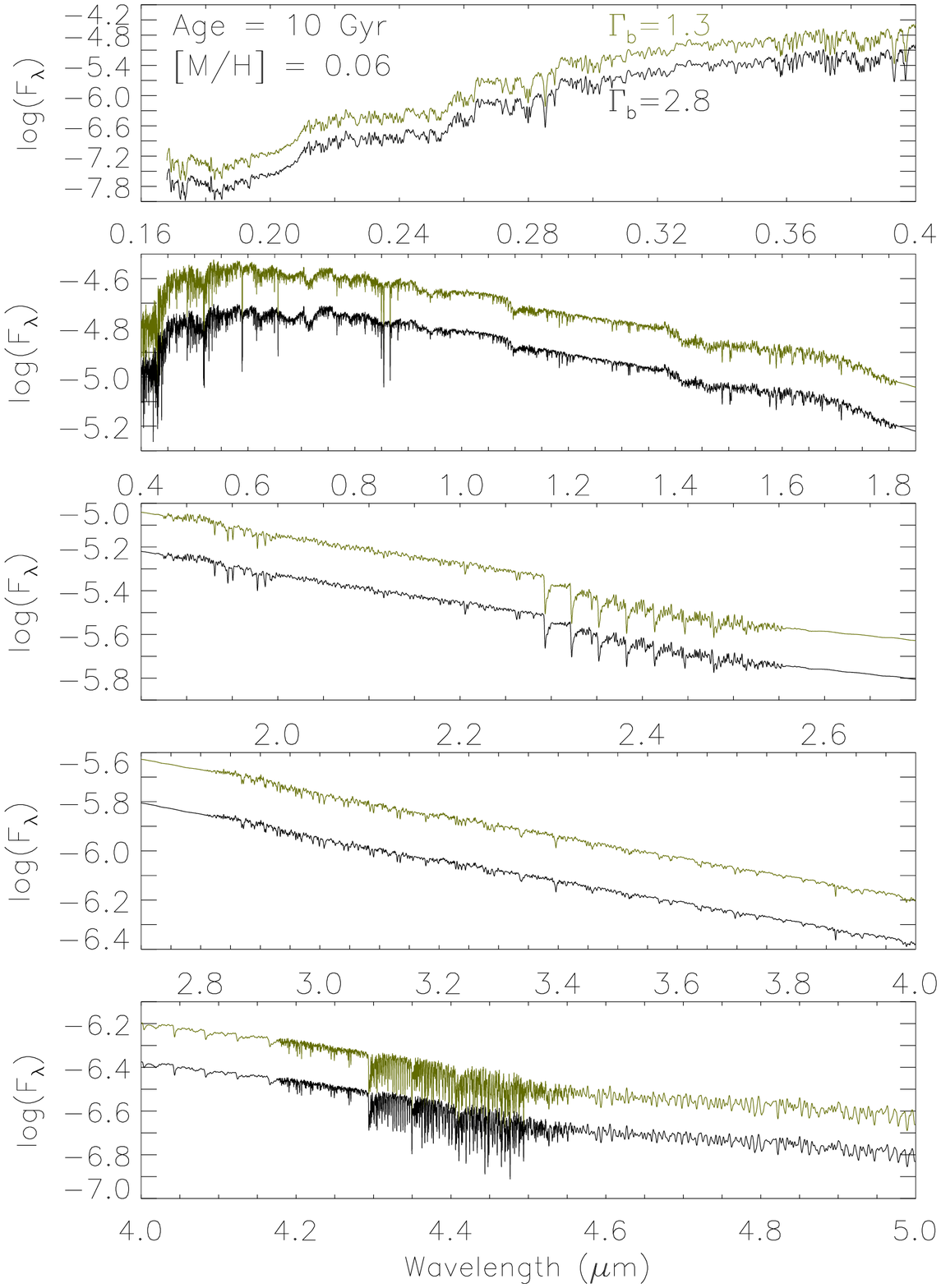}
   \caption{E-MILES SSP spectra of with age $12$\,Gyr, metallicity \Mh$=+0.06$
   and bimodal IMF with slope $\Gamma_{b}=1.3$ (olive green) and $\Gamma_{b}=2.8$
   (black). The model with $\Gamma_{b}=1.3$ has been shifted by
   $+\Delta\log{F_{\lambda}}=2.0$ in the first (upper) panel and by
   $+\Delta\log{F_{\lambda}}=1.3$ in all the other four panels for better
   visibility.}
   \label{fig:emiles_IMF}%
\end{figure*}

The resulting E-MILES spectra for two representative SSPs with age $12$\,Gyr and
with metallicity \Mh$=+0.06$ and \Mh$=-0.35$ are shown in
Fig.\,\ref{fig:emiles_12Gyr}. Similarly, in Fig.\,\ref{fig:emiles_2Gyr} we show
two representative models with age $2$\,Gyr. Finally, in
Fig.\,\ref{fig:emiles_IMF} we illustrate the effect of the IMF for an old model
of nearly solar metallicity. The selected IMF slopes can be considered
representative for low- and high-mass ETGs, following the results of
\citet{LaBarbera13}. All the SSP spectra plotted in these three figures can
be considered safe within the whole spectral range covered by the E-MILES
models, i.e. $\lambda\lambda$ $1680.2$--$49999.4$\,\AA.

The SSP models plotted in these three figures, including the youngest ones,
show many absorption features in the UV spectral range. In general these
features become stronger with increasing age and metallicity, as is the case for
most optical metallicity indicators. However metal line blanketing is
significantly larger in the UV than in the optical range. This makes it very
difficult to define reasonably well behaved index pseudocontinua bandpasses that
are needed to measure the strengths of the main absorption lines. In fact the
deepening and overlapping of metal lines surrounding some strong features migh
cause their strenths to decrease with increasing age/metallicity. This is the
case of, e.g., the Mg{\sc II} feature at $2800$\,\AA\ \citep{Smith91}. Moreover
this feature is part of a strong Mg lines complex that also includes the Mg{\sc
I} feature centered on $2852$\,\AA, which originates a prominent break at
$\sim2920$\,\AA\ and a rather wide spectral dip. This dip, which is
characteristic of cool stars and evolved stellar populations, is quantified by
the so called Mg wide index \citep{Fanelli90}, and can be used to study, e.g.,
higher redshift galaxies for which is not possible to obtain high S/N spectra.
Similarly, the break seen at $\sim2600$\,\AA\ is caused by a complex dominated by Fe
lines. As the spectral breaks bracketing these dips become more prominent with
increasing age and metallicity they can be used as indicators to constrain these
parameters. In fact \citet{Fanelli90} proposed specific indices composed by a
single pseudo-continuum and a feature bandpass to quantify these jumps, namely
$2609/2660$ and $2828/2921$. Note also that there are specific features that are
only noticeable in certain age/metallicity regimes. These include
pseudo-continua that may look as emission like features, such as, e.g. that at
$\sim2000$\,\AA, which is mostly seen in the more metal-rich model plotted in
Fig.\,\ref{fig:emiles_2Gyr}, as the surrounding absorption lines deepen with
increasing metallicity. Note also the trough seen blueward $2100$\,\AA\ that is
characteristic of evolved metal-rich stellar populations (see
Fig.\,\ref{fig:emiles_12Gyr}), such as massive elliptical galaxies. In
Section\,\ref{sec:model_lines} we describe in detail the behaviour of the most
prominent line indices in the UV spectral range covered by our models.

\section{Behaviour of the models in the UV spectral range}
\label{sec:behaviour}

In this section we focus on the behaviour of the colours, spectra and
line-strengths of our SSP models in the UV spectral range. We refer the reader
to the specific papers listed in Table\,\ref{tab:SEDproperties} for a detailed
description of the behaviour of our models in other wavelength ranges.

\subsection{Colours}
\label{sec:model_colours}

\begin{figure}
   \centering
  \includegraphics[width=0.49\textwidth]{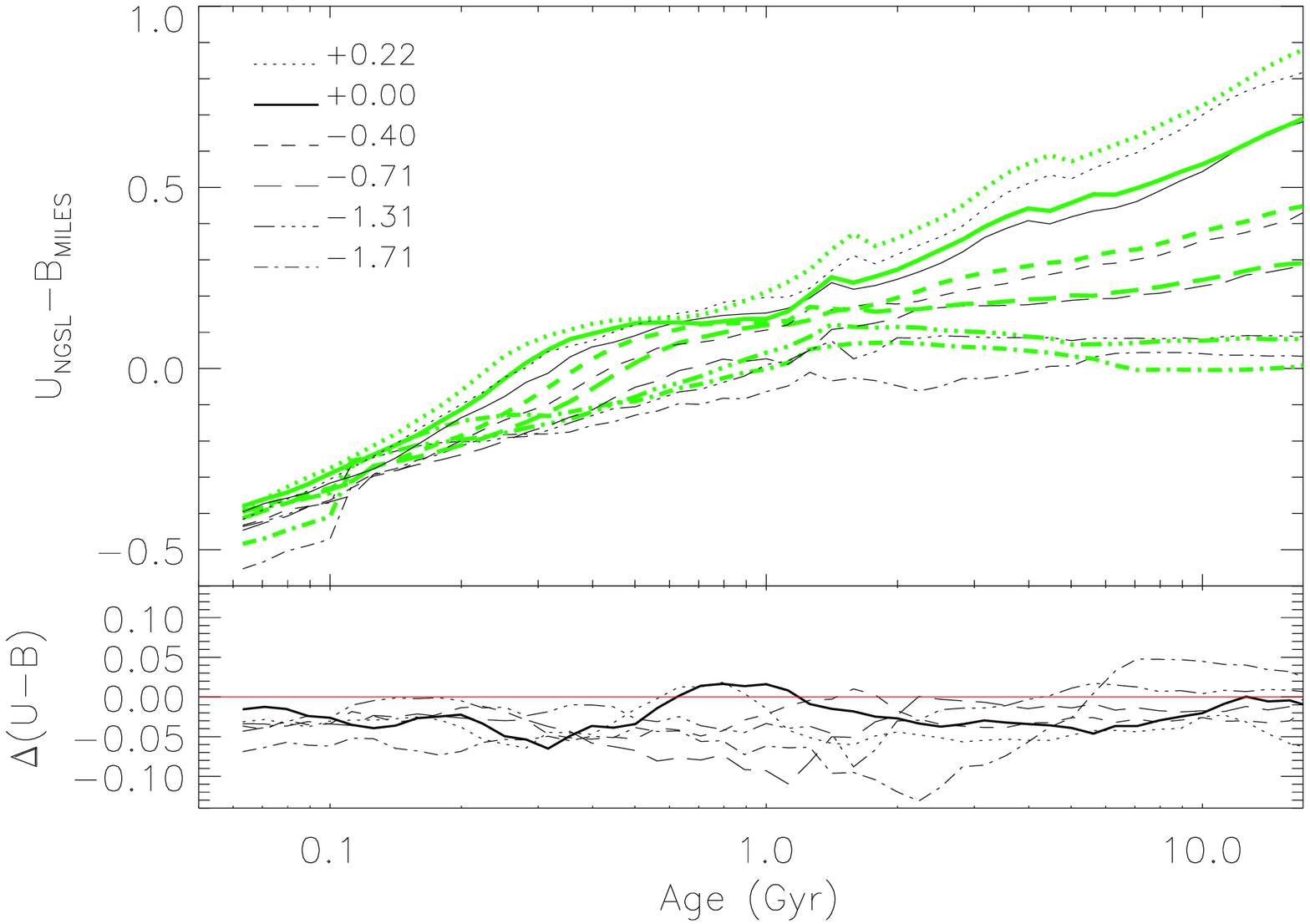}

   \caption{Time evolution of the U-B colour synthesised from our SSP model
   spectra (black lines), based on the Padova00 isochrones,  and from our
   photometric predictions (light green thick lines) based on extensive photometric stellar
   libraries (see the text for details). The metallicity values of the SSPs are
   represented with varying line types as quoted within the upper panel. The
   lower panel shows the residual colour, i.e. synthetic minus
   photometric colour. }

   \label{fig:UB}%
\end{figure}

The flux calibration of the resulting SSP model spectra can be tested by
comparing the synthetic U-B colour with our photometric predictions, which
were computed on the basis of the extensive photometric stellar libraries
described in Section\,\ref{sec:models}. The U filter covers the overlapping
region between the NGSL and MILES at $\sim3540$\,\AA, whereas the B filter is
entirely measured within the MILES spectral range. The synthetic U$-$B colour
is calculated using the \citet{BuserKurucz78} filter definitions and the Vega
magnitude system (see \citealt{FalconBarroso11} for further details). The
photometrically predicted colour is from \citet{Vazdekis96} as updated in
\citet{MILESIII}, and is based on empirical relations between colours and
stellar parameters \citep{Alonso96,Alonso99}. The residual U$-$B colour shown
in Fig.\,\ref{fig:UB} is within typical zero-point errors ($\sim0.02$\,mag)
for the old models with solar metallicity. In this case the largest residuals
are found for the intermediate age regime. However for the most metal-poor
models ($\Mh=-1.71$) we obtain significantly larger residuals. It is worth
recalling that, as discussed in \citet{MIUSCATI} and \citet{MIUSCATII},
the synthetic U magnitude is very sensitive to the definition of the U filter
and the empirical colour-temperature relations used for the U$-$B colour are
not as homogenous as in the other pass-bands, which makes it more difficult
to match. We conclude that the inclusion of the NGSL model spectra has
allowed us to obtain a significantly better agreement between the synthetic
and photometric colours than what was achieved with the MIUSCAT model spectra
alone (see Fig.\,8 of \citealt{MIUSCATI}).

\begin{figure}
   \centering
  \includegraphics[width=0.49\textwidth]{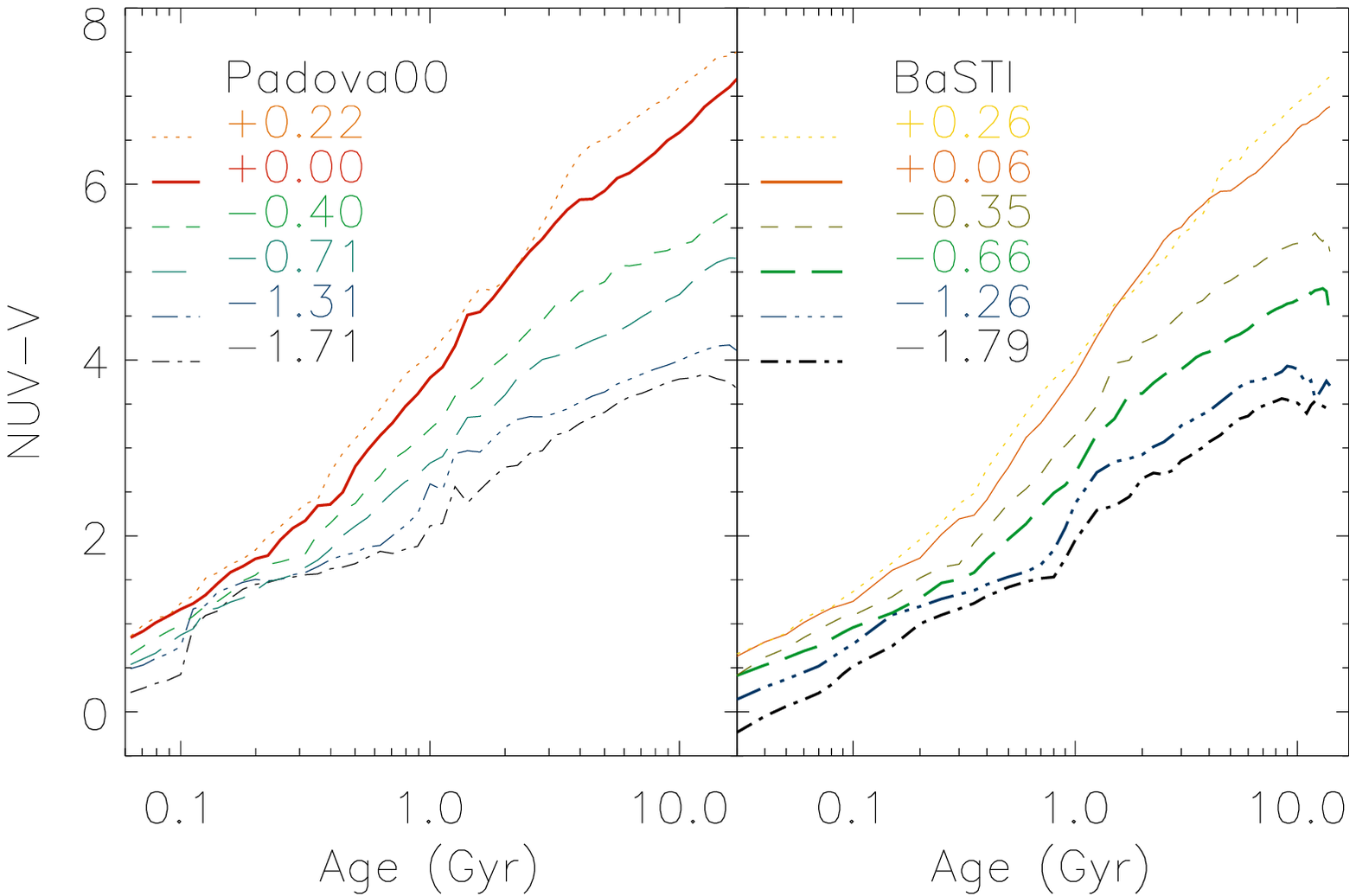}

   \caption{Evolution with time of the NUV-V colour synthesised from our SSP
   model based on Padova00 (left-hand panel) and BaSTI (right-hand panel)
   isochrones. The metallicities are represented with varying colour and
   line types as quoted within the panels.}

   \label{fig:NUVV}%
\end{figure}

The spectral range covered by our models allows us to measure the GALEX NUV
magnitude. Figure \ref{fig:NUVV} shows the theoretical predictions for the
NUV-V colour synthesized from our models, both based on the Padova00 and BaSTI
isochrones, as a function of age and metallicity. The NUV-V colour is given in
the AB system, using the NUV GALEX transmission curve and the V broad-band
filter of \citet{BuserKurucz78}. Similarly to the optical colours the NUV-V
reddens with increasing metallicity. Howerver in comparison to the U-B colour
(see Fig.\,\ref{fig:UB}) and the other optical colours, which tend to increase
more modestly for ages above $1$\,Gyr, the NUV-V colour shows a steady and
sharp increase with increasing age. Such a behaviour can be attributed to the
strong contribution of the turnoff stars and the low contribution of the RGB
phase in the NUV spectral range. For sub-solar and very metal-poor
populations, the BaSTI-based models show a drop in colour for ages larger than
10\,Gyr. This effect is more modest in the Padova00-based models as the
Horizontal-Branch is cooler for such old ages.

\begin{figure}
   \centering
  \includegraphics[width=0.49\textwidth]{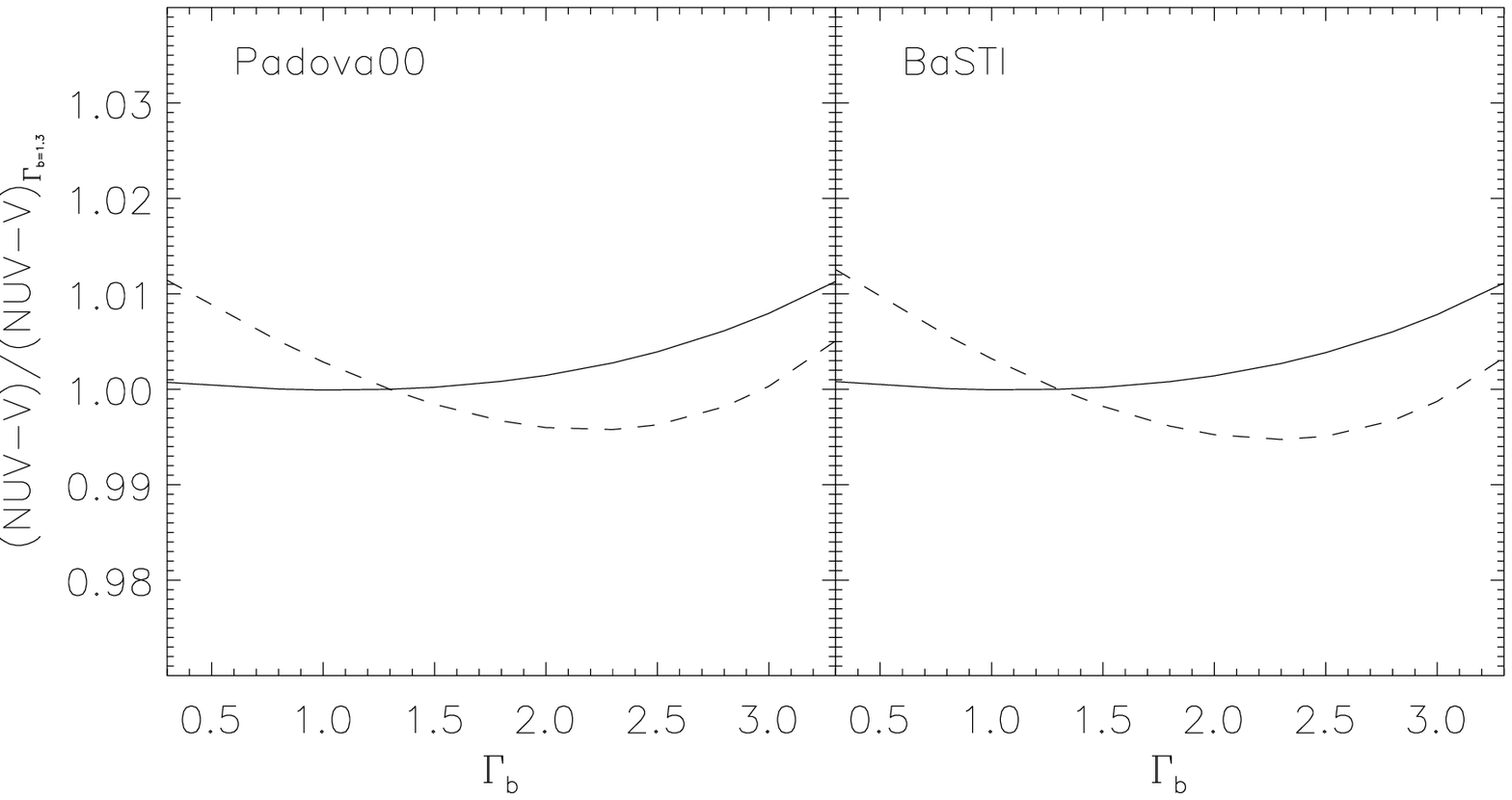}

   \caption{Dependence of the NUV-V colour on the bimodal IMF slope $\Gamma_b$ 
 for models of 10\,Gyr (solid line) and 2\,Gyr (dashed line) and solar
 metallicity. The obtained difference in colour with respect to the model with
 standard IMF slope, i.e. $\Gamma_{b}=1.3$, is shown as a function of
 $\Gamma_b$, for models based on Padova00 (left panel) and BaSTI (right panel)
 isochrones.}

   \label{fig:NUVVimf}%
\end{figure}

The dependence of the NUV-V colour on the IMF slope is shown in Figure
\ref{fig:NUVVimf} for the bimodal IMF type. The figure shows the colour residual
obtained for models with varying slope when compared to the reference standard
model with $\Gamma_{b}=1.3$. Solar metallicity models and two representative SSP
ages are shown. The impact of the IMF on the UV colour appears negligible, i.e.
well below typical zero-point errors, for the two ages. We conclude that the IMF
has little effects in this spectral range. 

\subsection{Spectra}
\label{sec:model_spectra}

\begin{figure}
   \centering
  \includegraphics[width=0.49\textwidth]{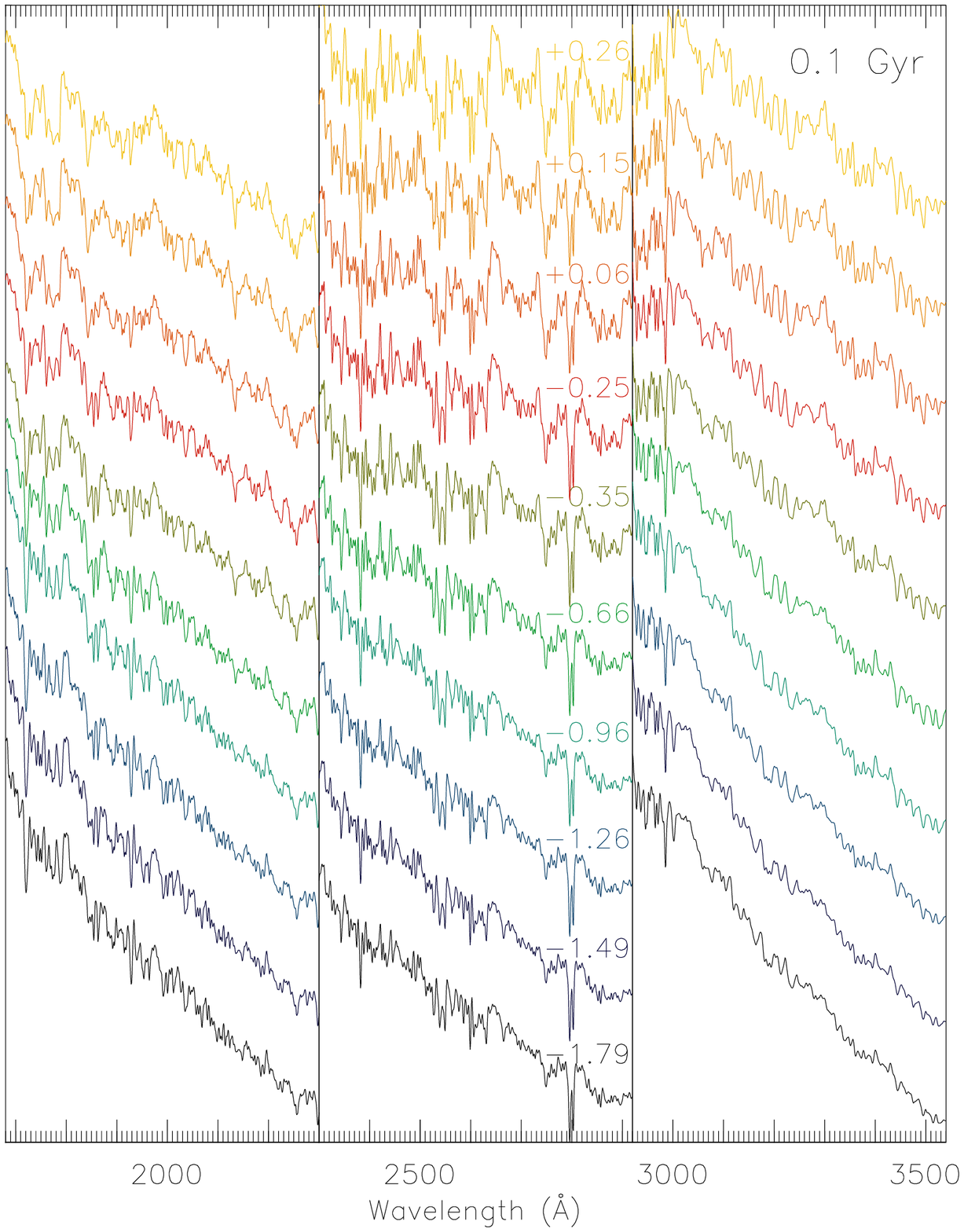}
   \caption{SSP spectra with age $0.1$\,Gyr and decreasing metallicity, from top
to bottom as marked within the second window. All the models
were computed with BaSTI isochrones and Kroupa Universal IMF. Note that the
sharp spectrum shape of these models does not allow us to appreciate the details
in the spectra if we do not split them in various windows.}
   \label{fig:models01}%
\end{figure}

We plot in Fig.\,\ref{fig:models01} the UV range of a set of representative
SSP spectra with young age ($0.1$\,Gyr) and varying metallicity. Note that the
strong flux variation within this wavelength range requires plotting the model
spectra in various windows to be able to appreciate the features. The flux
increases toward the shortest wavelengths. This effect is more pronounced for
the more metal-poor stellar populations. Such an effect is attributed to the
contribution of the turnoff stars, which become hotter with decreasing
metallicity.

\begin{figure}
   \centering
  \includegraphics[width=0.49\textwidth]{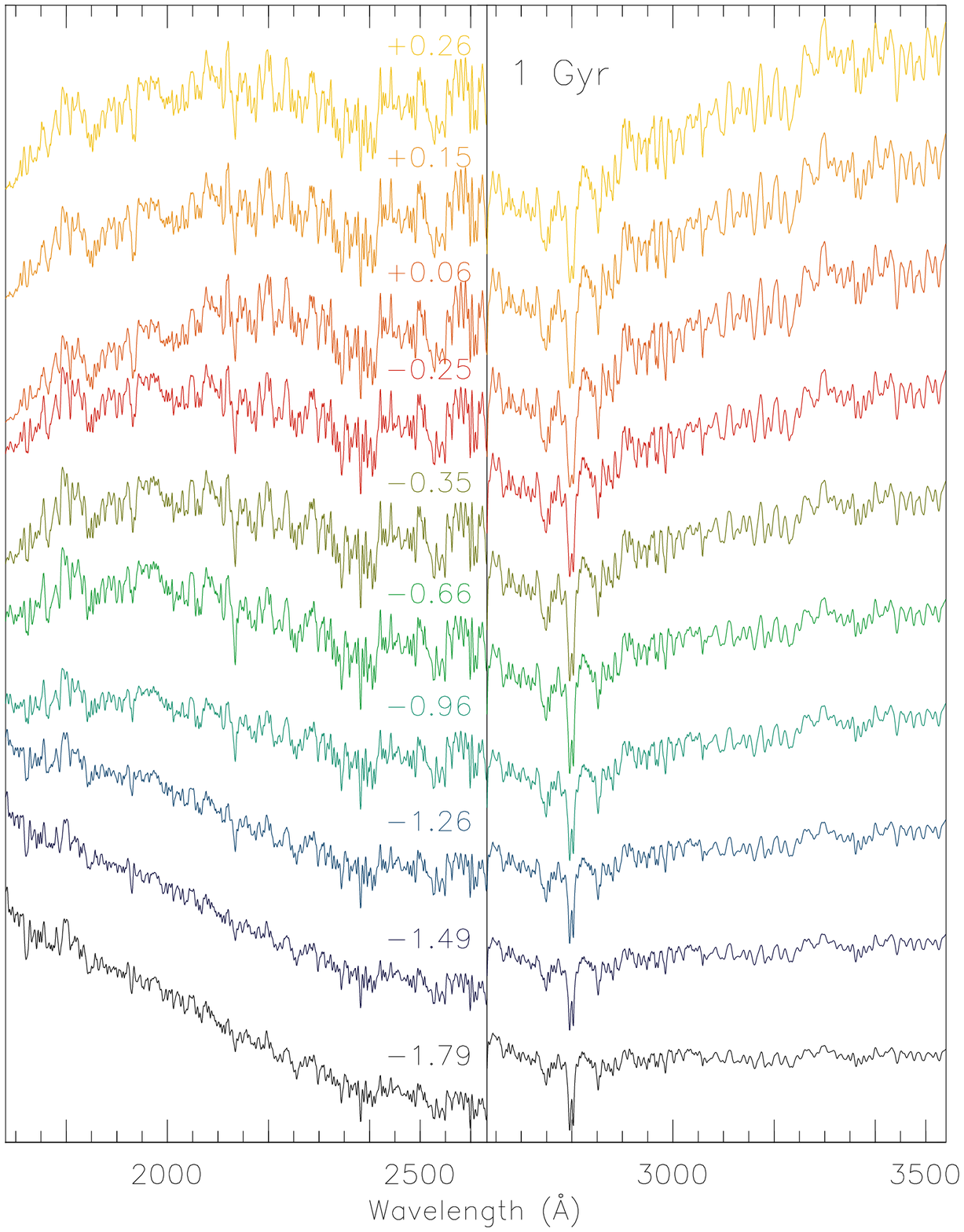}
   \caption{Same as in Fig.\,\ref{fig:models01} but for models of $1$\,Gyr. }
   \label{fig:models1}%
\end{figure}

In Fig.\,\ref{fig:models1} we show SSP spectra of models with intermediate
age ($1$\,Gyr) and varying metallicity. In this case two windows are
sufficient to visualize the details of the spectra as the spectrum shape is
flatter than in the models shown in Fig.\,\ref{fig:models01}. Note that for
this age the models with the lowest metallicities still show an increasing
flux for the bluest wavelengths, whereas the most metal-rich SSPs become
significantly redder. 

\begin{figure}
   \centering
  \includegraphics[width=0.49\textwidth]{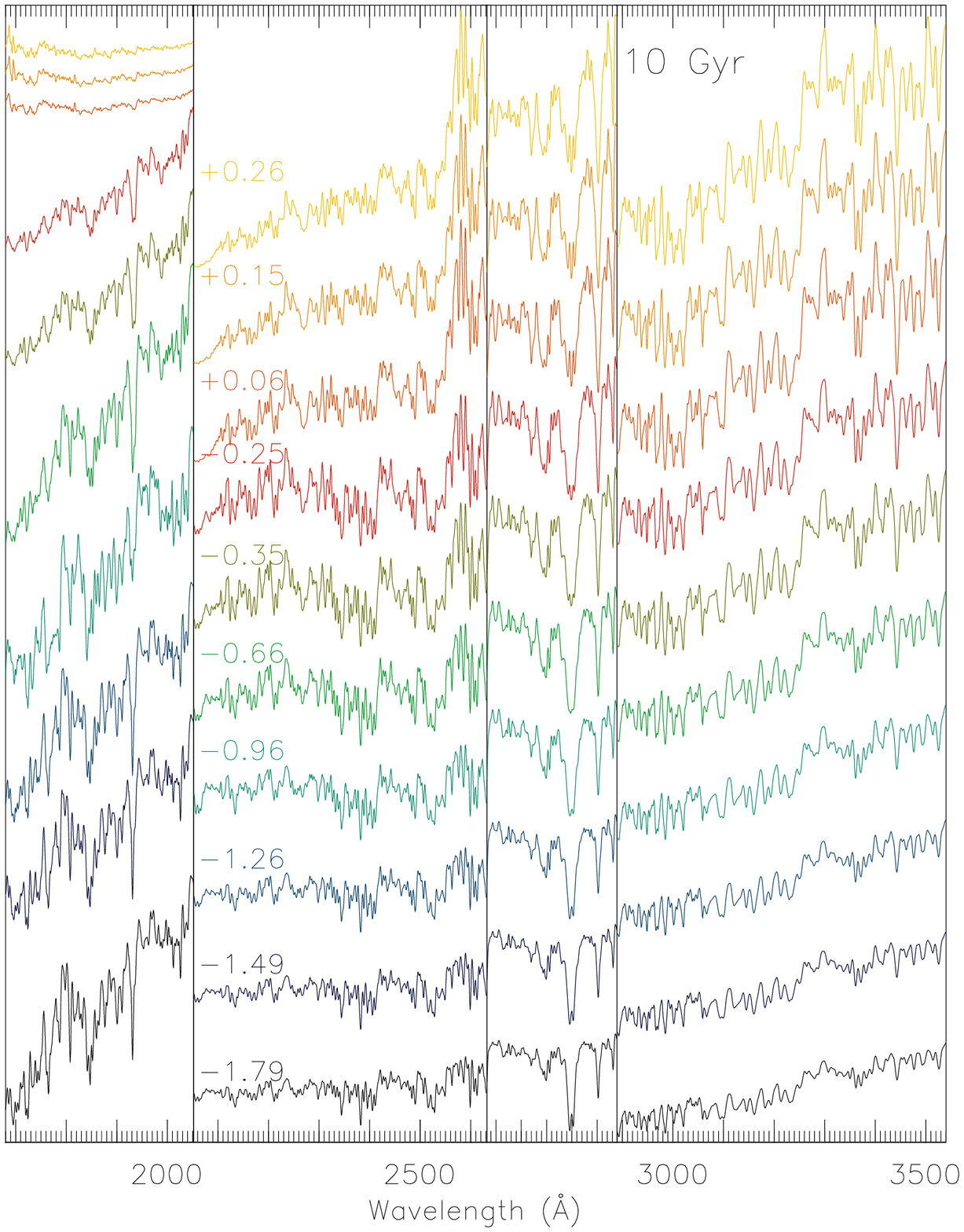}
   \caption{Same as in Fig.\,\ref{fig:models01} and Fig.\,\ref{fig:models1} but for models of $10$\,Gyr. }
   \label{fig:models10}%
\end{figure}

Finally, in Fig.\,\ref{fig:models10} we show the old ($10$\,Gyr) models for
the same metallicities. In contrast to the younger models, in this case the
flux increases with increasing wavelength in all the plotted SSP spectra. The
sharp spectrum shape forced us to split this spectral range in four windows to
be able to see the features, which are much stronger than in the youngest
models. It is remarkable that some features, such as the Mg absorption around
$2800$\,\AA, become stronger with decreasing metallicity. This is however not
the case for the neighbouring Mg absorption at $2852$\,\AA. We discuss this
issue in more detail in Section\,\ref{sec:model_lines}.

\subsection{Line-strengths}
\label{sec:model_lines}

The UV model extension based on the NGSL allows us to work with line indices
that are not covered by the SSP spectra computed with MILES.
Table\,\ref{tab:indices} lists the selected indices in this spectral range.
Although our models allow us to optimize these indices for studying galaxy
spectra, we adopt here the definitions from various sources, mostly from
\citet{IUEII}. All the selected indices are defined by a feature bandpass and
a pseudo-continuum at either side of the feature. In most cases the index
definitions do not extend beyond $\sim200$\,\AA, from the bluest wavelength of
the blue pseudocontinuum to the reddest wavelength of the red pseudocontinuum,
with the notable exception of the Mg wide index, which covers 660\,\AA. The
table also lists the main elements contributing to these features. We see that
these lines are composed by different elements, being strongly blended in many
cases. Note that the dominating elements might be at varying ionization stages (e.g.,
Mg\,$_{\rm \sc II}$\,2800 and Mg\,$_{\rm \sc I}$\,2852). Note also that the
prominent Mg\,2800 feature might be affected by chromospheric emission
filling-in.  

\begin{table*}
\centering
\caption{\label{tab:indices} UV Line-strength indices.}
\begin{tabular}{lcccll}
Index & Blue Passband & Index Passband & Red Passband & Contributions & Reference \\
\hline                   
BL\,1719 & 1685 1705 & 1709 1729 & 1803 1823 & N,S,Al     &1,2\\
BL\,1853 & 1803 1823 & 1838 1868 & 1885 1915 & Al,Fe      &1,2\\
Fe\,2332 & 2285 2325 & 2333 2359 & 2432 2458 & Fe,Co,Ni   &1,3\\
Fe\,2402 & 2285 2325 & 2382 2422 & 2432 2458 & Fe,Co      &1,2,3\\ 
BL\,2538 & 2432 2458 & 2520 2556 & 2562 2588 & Fe,Mg,Cr,Ni&1,2,3\\
Fe\,2609 & 2562 2588 & 2596 2622 & 2647 2673 & Fe,Mn      &1,2,3\\
BL\,2720 & 2647 2673 & 2713 2733 & 2762 2782 & Fe,Cr      &1,3\\
BL\,2740 & 2647 2673 & 2736 2762 & 2762 2782 & Fe,Cr      &1,3\\
Mg\,2800 & 2762 2782 & 2784 2814 & 2818 2838 & Mg,Fe,Mn   &1,2,3\\
Mg\,2852 & 2818 2838 & 2839 2865 & 2906 2936 & Mg,Fe,Cr   &1,2,3\\
Mg wide  & 2470 2670 & 2670 2870 & 2930 3130 & Mg,Fe,Cr   &1,2,3\\
Fe\,3000 & 2906 2936 & 2965 3025 & 3031 3051 & Fe,Cr,Ni   &1,2,3\\
BL\,3096 & 3031 3051 & 3086 3106 & 3115 3155 & Fe,Ni,Mg,Al&1,2,3\\
NH\,3360 & 3320 3350 & 3350 3400 & 3415 3435 & N,Mg,Fe,Ni &4\\
NH\,3375 & 3342 3352 & 3350 3400 & 3415 3435 & N,Ti,Ni    &5\\
Mg\,3334 & 3310 3320 & 3328 3340 & 3342 3355 & Mg,O       &5\\
\hline
\\
\multicolumn{6}{l}{1 \citet{Fanelli90}}\\
\multicolumn{6}{l}{2 \citet{Maraston09}}\\
\multicolumn{6}{l}{3 \citet{Chavez07}}\\
\multicolumn{6}{l}{4 \citet{DavidgeClark94}}\\
\multicolumn{6}{l}{5 \citet{Serven11}}\\
\end{tabular}
\end{table*}

\begin{figure*}
   \centering
  \includegraphics[width=0.99\textwidth]{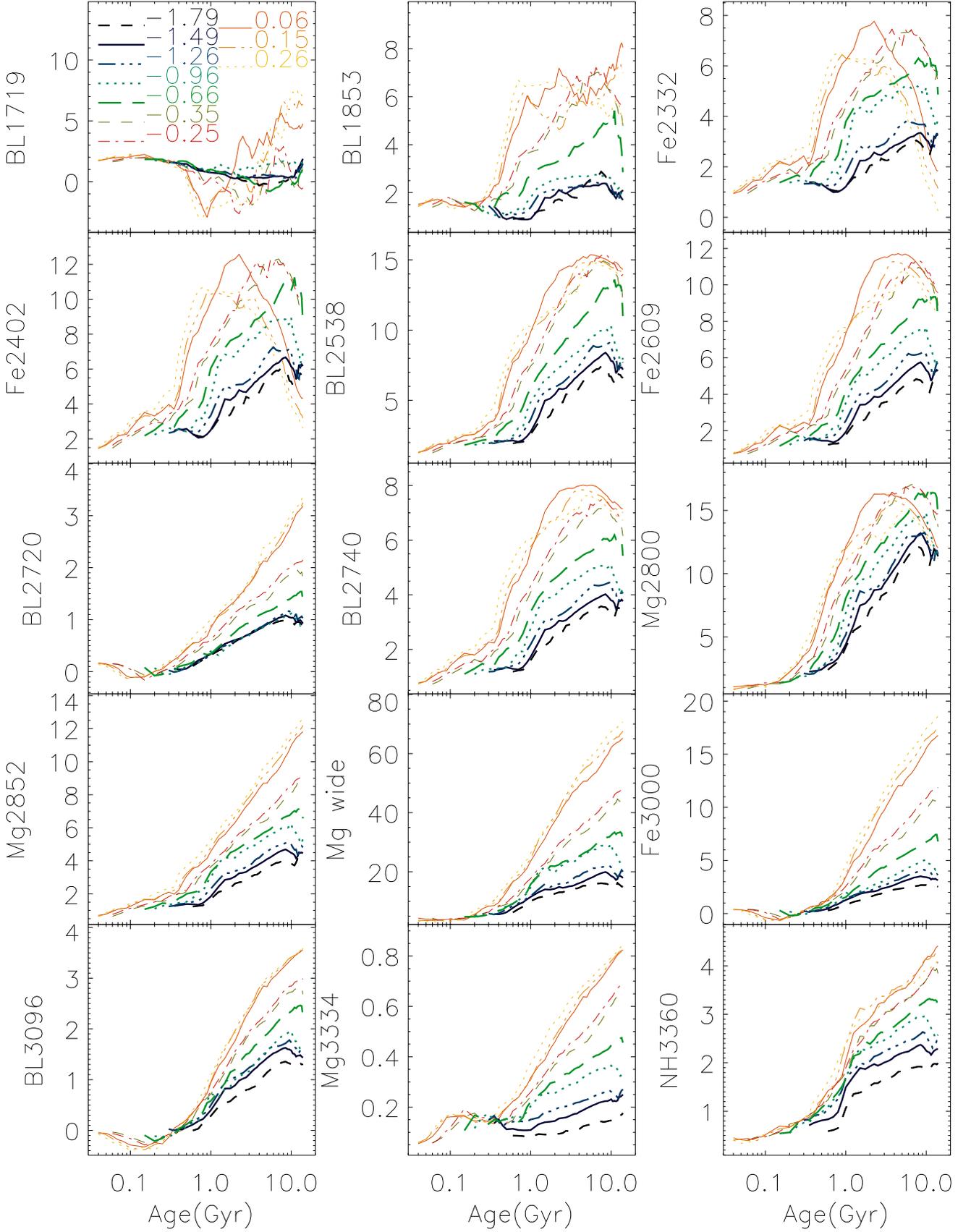}
   \caption{Behaviour of the UV line-strengths as a function of age and
   metallicity as quoted in the first panel. The SSP spectra, which were
   computed with the BaSTI isochrones adopting a Kroupa Universal IMF, were
   smoothed to 5\,\AA\ (FWHM) ($\sigma\sim210$\,\kms\ at $3000$\,\AA) to match the
   LIS-5.0\,\AA\ system defined in \citet{MILESIII}.}
   \label{fig:UVlines_Teramo}%
\end{figure*}

\begin{figure*}
   \centering
  \includegraphics[width=0.99\textwidth]{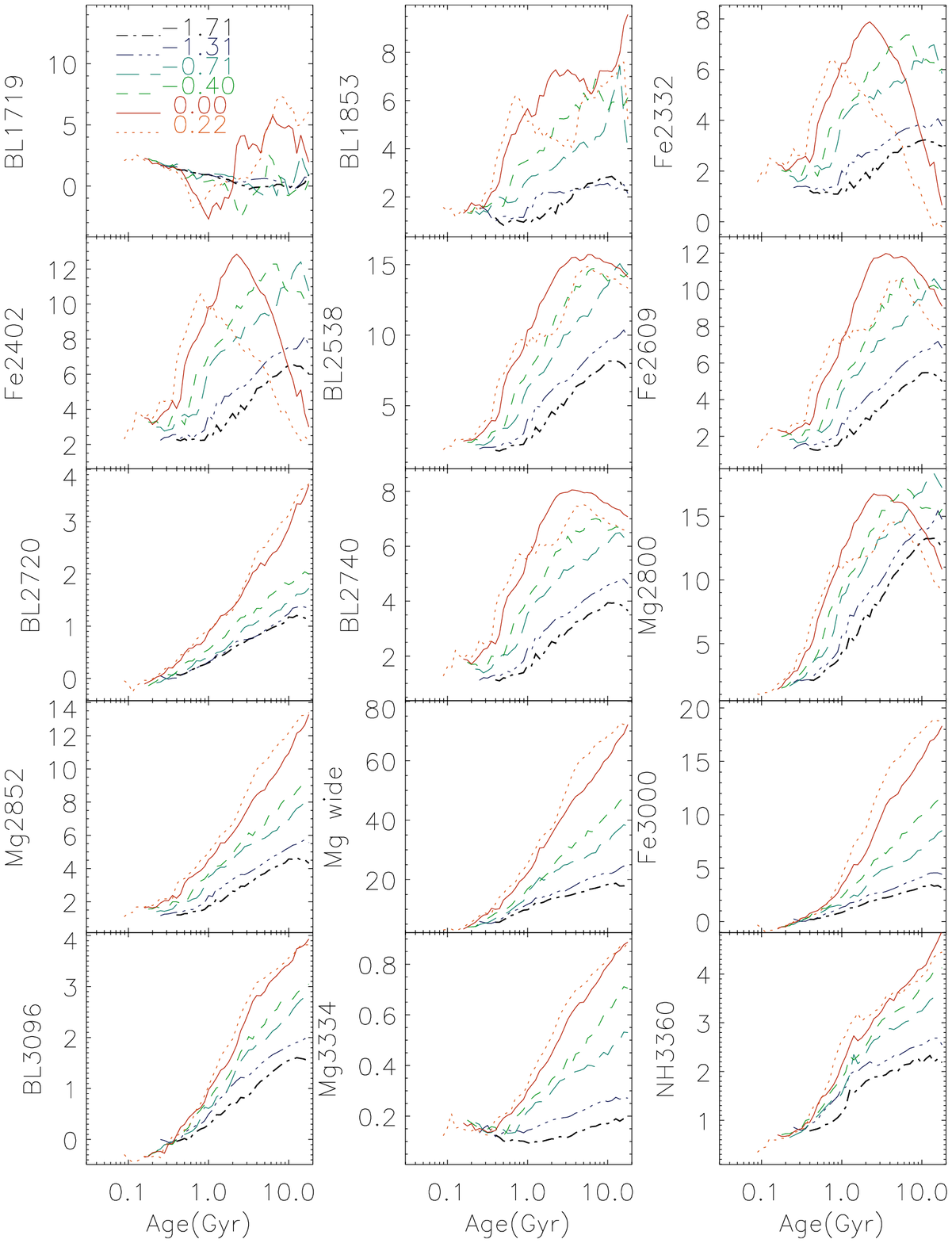}
   \caption{
  Same as Fig. \ref{fig:UVlines_Teramo}, but for models computed with
  the Padova00 isochrones.}
   \label{fig:UVlines_Padova}%
\end{figure*}

In Fig.\,\ref{fig:UVlines_Teramo} and Fig.\,\ref{fig:UVlines_Padova} we show
the behaviour of most of the UV line indices listed in
Table\,\ref{tab:indices} as a function of age for the SSP models computed with
the BaSTI and Padova00 isochrones, respectively. The SSP spectra were smoothed
to 5\,\AA\ (FWHM) ($\sigma\sim210$\,\kms\ at $3000$\,\AA) to match the
LIS-5.0\,\AA\ system defined in \citet{MILESIII}. Note that this is the
nominal resolution of the models in the spectral range $\lambda\lambda$
$3060.8-3541.4$\,\AA\ (see Table\,\ref{tab:SEDproperties}). The indices are
ordered by increasing wavelength from the top-left to the bottom-right
panels. 

The strength of the reddest index, i.e. NH\,3360, increases with increasing
age and increasing metallicity, but shows a relatively more modest increase
with age for ages larger than $\sim1$\,Gyr. Such a behaviour is characteristic
of many indices in the optical range (e.g. \citealt{MILESIII}) and is related
to the RGB phase. In the NIR range many indices tend to flatten even more (in
some cases decrease) with increasing age (e.g. \citealt{Roeck16}). Note that
for sub-solar metallicity models and ages above $\sim10$\,Gyr the strength of
the index tends to decrease due to the Horizontal-Branch bluening (more
notorious in the BaSTI based models), as it happened to the NUV-V colour (see
Section\,\ref{sec:model_colours}).  

The observed line-strength flattening of the NH\,3360 index for ages above
$\sim1$\,Gyr is not seen for other indices with feature bandpasses redder
than $3000$\,\AA\ (Mg\,3334, BL\,3096, and Fe\,3000). This effect might be
explained by the fast evolution of the temperature of the turnoff stars,
which are more influential at these wavelengths (e.g. \citealt{Chavez07}),
whereas the RGB stars decrease their weight. The behaviour of these indices
is therefore more similar to that of the NUV-V colour (see
Section\,\ref{sec:model_colours}). This behaviour also applies to the Mg wide
and Mg\,2852 indices, which show a steady increase with increasing age. 

Most of the indices centered within the range $\lambda\lambda$ $2300-2900$\,\AA\
show a very peculiar behaviour, as the index strengths of the more metal-rich
stellar populations peak at intermediate ages. Moreover, for the Mg\,2800,
Fe\,2402 and Fe\,2332, the index values corresponding to the metal-poor SSPs can
be larger than those of the metal-rich SSPs of very old ages. This result is in
agreement with the models of \citet{BC03}, based on theoretical stellar spectra,
as shown by \citet{Daddi05} (see their Fig.\,2).  This is a remarkable result,
as our models predict a larger, e.g., Mg\,2800 index values for the Milky-Way
globular clusters with metallicities around \Mh$\sim-0.7$ than in giant
elliptical galaxies. We discuss this issue in more detail in
Section\,\ref{sec:GCslines} and Section\,\ref{sec:ETGslines}. It is worth
noticing that the predicted strengths for the Mg\,2800 index corresponding to
the models with the youngest ages might be partially biased by interstellar
absorption affecting some of the hot stars present in the NGSL (see
Section\,\ref{sec:NGSL}). Note that in this respect the NGSL is not particularly
different from the IUE stellar library employed by the models of
\citet{Maraston09}. Therefore, judging from their Figure 10 where these authors
show their predictions based on both the IUE stellar library and on theoretical
spectra, we conclude that the net effect is smaller than $\sim1$\,\AA.
Furthermore, as in our case we decreased the relative weight of these stars when
computing our SSP spectra (see Section\,\ref{sec:NGSL}), this effect is even
smaller. Caveats should be taken into account by potential users of our models
when applying them to this feature in such a young age regime. 

Finally, the behaviour of the blended features BL\,1719 and BL\,1853, which
fall within the bluest end of our spectral range coverage, resemble in part
that of NH\,3360. However for the age regime below $\sim1$\,Gyr these indices
decrease slightly with increasing age. This happens until reaching a certain
age value, which is larger with decreasing metallicity.       

\begin{figure}
   \centering
  \includegraphics[width=0.49\textwidth]{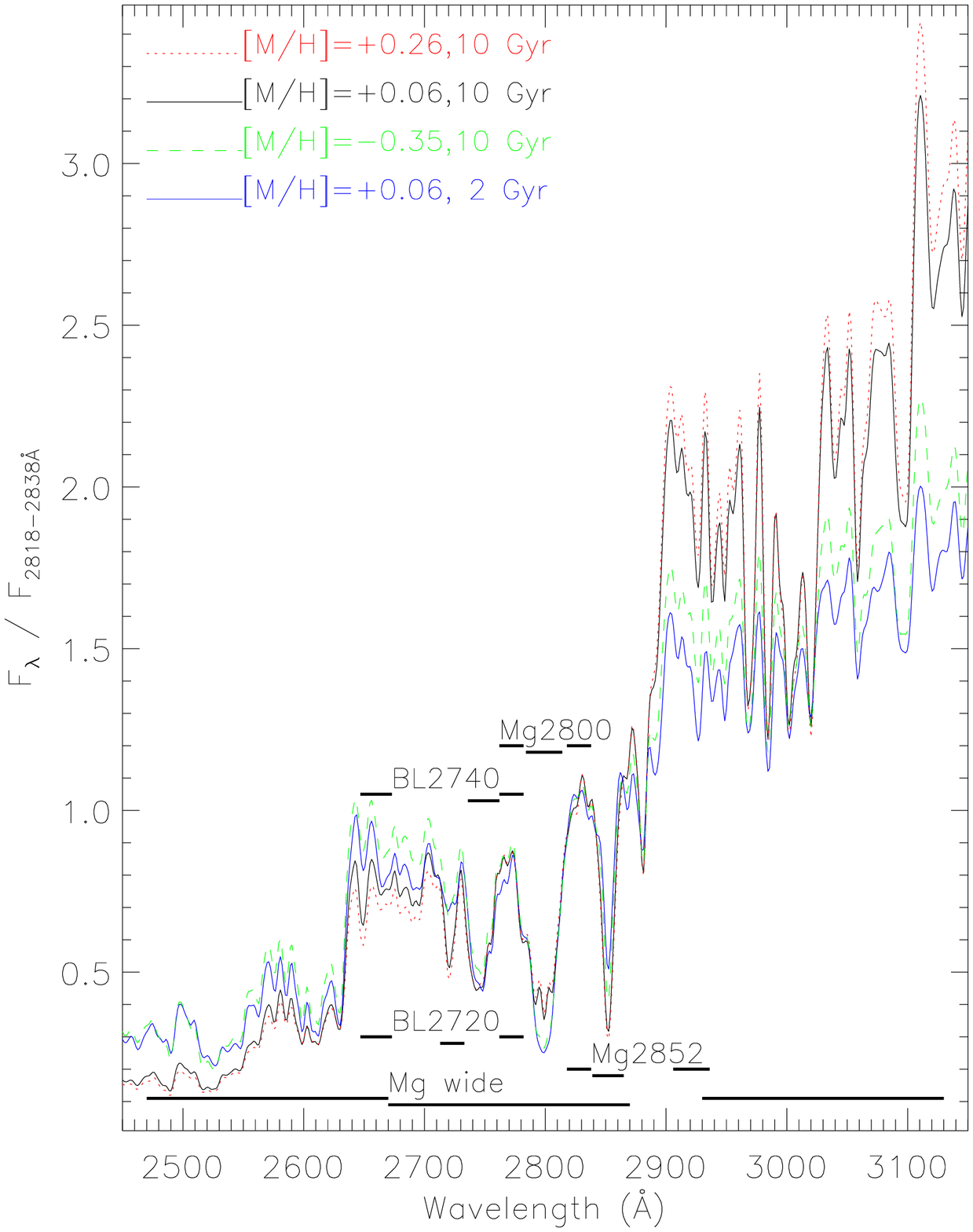}
   \caption{Spectral region around the Mg\,2800 feature with a number of SSP
   spectra of various ages and metallicities, as quoted within the panel,
   overplotted. All the spectra were normalized at the red pseudocontinuum of
   the Mg\,2800 index. The feature and psedocontinua bandpasses corresponding to
   various indices in this spectral region are placed at a relatively slightly
   lower and higher flux level, respectively. The strengths of the indices
   marked above the spectra (i.e., BL\,2740 and Mg\,2800) peak at ages around
   $\sim2$\,Gyr for metal-rich stellar populations, in contrast to the ones
   marked below the spectra. See for details Figs.\,\ref{fig:UVlines_Teramo} and
   \ref{fig:UVlines_Padova} and the text.}
   \label{fig:Mg2800}%
\end{figure}

In Fig.\,\ref{fig:Mg2800} we zoom in the spectral region around the Mg\,2800
feature. The selected window covers the definition of the Mg wide index.
Overplotted are various SSP models with $10$\,Gyr and different
metallicities. Also plotted is a model of $2$\,Gyr and \Mh$=+0.06$ for which
the strengths of the  Mg\,2800 and BL\,2740 indices reach their maximum
values. In contrast to these two indices, whose pseudocontinua and feature
bandpasses are shown above the SSP spectra, the three indices shown below the
spectra, namely Mg wide, BL\,2720 and Mg\,2852, increase their strength with
increasing metallicity and increasing age.  These spectra show that the Mg
wide index measures a strong bump in the spectrum, which includes the two
blends at 2720\,\AA\ and 2740\,\AA\ and the two Mg-dominated absorption
features. As the metallicity and/or age increases the bump becomes more
prominent and so the strength of this index. Similarly the two narrow indices
BL\,2720 and Mg\,2852 show a very strong sensitivity to the metallicity and
to the age. Furthermore, the sensitivity of the Mg\,2852 index to these
parameters is emphasized by the location of its red pseudocontinuum, redward
$\sim2900$\,\AA, just on the top of a prominent flux jump, which is more
pronounced for increasing metallicity/age with respect to the flux of its
blue pseudocontinuum that sits within the bump.  In contrast to Mg\,2852 the
Mg\,2800 and BL\,2740 indices do not show such sensitivity to the
metallicity. As the two pseudocontinua are placed within the bump,  a deeper
bump leads to smaller fluxes, which plays against strengthening the measured
index values. Fig.\,\ref{fig:Mg2800} shows that the behaviour of these
indices as a function of metallicity and age does not solely depend on the
depth of their absorption lines. 

\section{Comparison with other models}
\label{sec:comparison}

\begin{figure}
   \centering
  \includegraphics[width=0.49\textwidth]{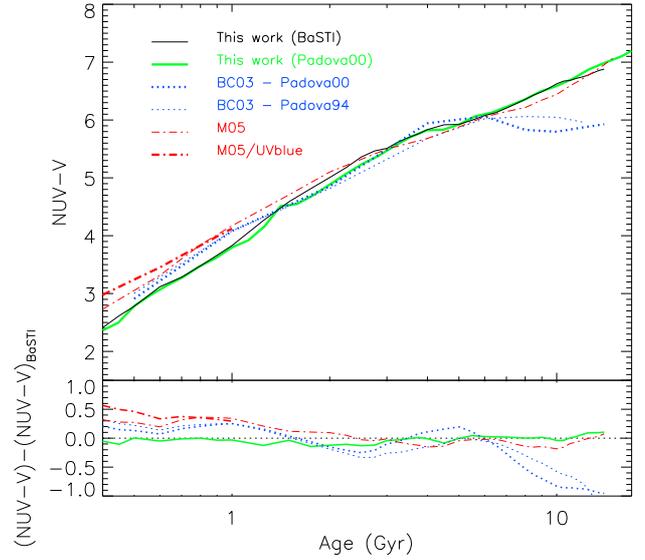}
   \caption{Comparison of NUV-V colour predictions from different authors, for
   solar metallicity SSPs and Kroupa Universal IMF. Black thin and green thick
   solid lines indicate our models, based on BaSTI and Padova00 isochrones,
   respectively. The blue lines represent the \citet{BC03},  based on
   \citet{Padova94} (thin dotted) and \citet{Padova00} (thick dotted)
   isochrones, and the red thin dot-dashed line represents the Maraston (2005)
   models. The thick red dot-dashed line shows a version of the latter models
   for ages smaller than 1\,Gyr, where the NUV magnitude is computed with the
   UVBLUE theoretical stellar spectral library \citep{UVBLUE}. The lower panel
   shows the residual colours with respect to our BaSTI-based models.}
   \label{fig:othercolours}%
\end{figure}

In this section we compare our models with other predictions found in the
literature. Fig.\,\ref{fig:othercolours} shows the behaviour of the NUV-V
colour as a function of age for various SSP models. The figure shows the
colours measured on our SSP spectra computed with the BaSTI and Padova00
isochrones, together with the colours of the models of \citet{BC03} for two
different versions of the Padova isochrones \citep{Padova94,Padova00} and
those of the models of \citet{Maraston05}. For the latter models, we also
show a version that makes use of a high resolution theeoretical library
(UVBLUE: \citealt{UVBLUE}) to obtain the NUV magnitudes for ages smaller than
$1$\,Gyr and solar metallicity. In all cases the models have solar metallicity.
The comparison between our models, which are virtually similar for all the plotted
ages,  strongly suggests that the differences, mostly in the turnoff, have a
minor secondary effect on the derived NUV magnitude. In comparison to the
other authors our predictions provide the reddest colours for the oldest
stellar populations and the bluest ones for the youngest stellar populations.
Overall, we obtain a good agreement with the \citet{Maraston05} models for the
whole age range shown in the figure. The young models of Maraston that
are based on the UVBLUE theoretical library provide the worst agreement with
our predictions. However the most significant disagreement is found when
comparing either version of the \citet{BC03} for the oldest stellar
populations. Their colours are around 1\,magnitude bluer than all the other
models shown here. This discrepancy can be attributed to the implementation
of planetary nebulae spectra in the \citet{BC03} models, which make their SSP
spectra particularly bluer at shorter wavelengths (private communication by
G. Bruzual).

\begin{figure}
   \centering
  \includegraphics[width=0.49\textwidth]{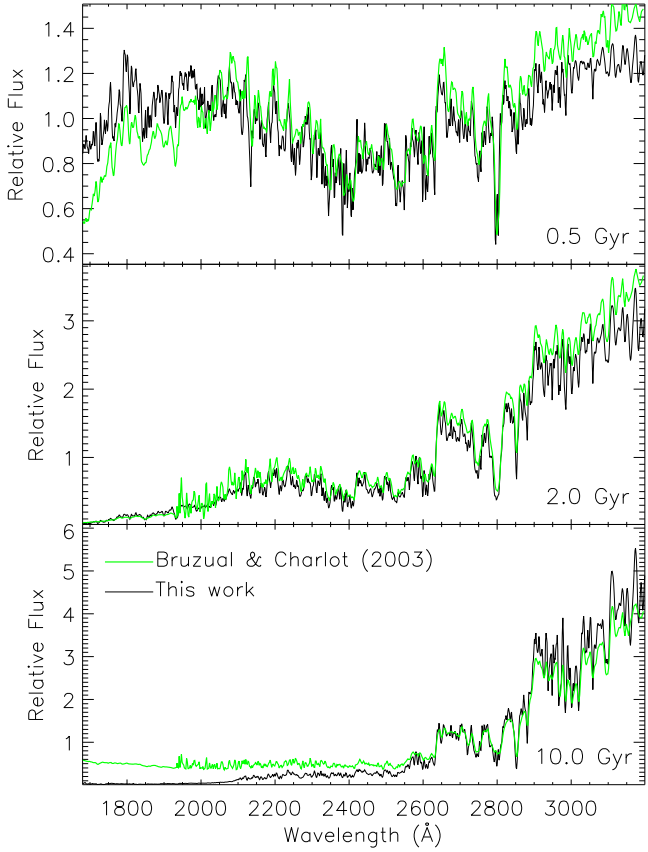}

   \caption{Comparison of various SSP spectra (black thin line) of varying
   ages ($0.5$, $2$ and $10$\,Gyr from the top to the bottom panel,
   respectively) with those of \citet{BC03} (green thick line). All the
   models have solar metallicity and are computed adopting a Kroupa Universal
   IMF.}

   \label{fig:bruzualspectra}%
\end{figure}

In Fig.\,\ref{fig:bruzualspectra} we compare our SSP spectra with those of 
\citet{BC03} based on the IUE stellar library for various representative ages,
namely $0.5, 2$ and $10$\,Gyr. Overall the models are in agreement for the young and
the intermediate age regimes, although there is a residual in the colour. The
main difference is found for the models of $10$\,Gyr, as expected from the results
obtained in Fig.\,\ref{fig:othercolours}. This discrepancy is more notorious at
the bluest wavelengths, where the effects of the planetary nebulae spectra
implemented in the \citet{BC03} models are larger. 

\section{Applications}
\label{sec:applications}
\subsection{Colours}

In order to assess the quality of our models, we compare the predictions for broad-band colours with data from globular
clusters and nearby galaxies.

\subsubsection{Globular clusters}

\begin{figure}
   \centering
  \includegraphics[width=0.49\textwidth]{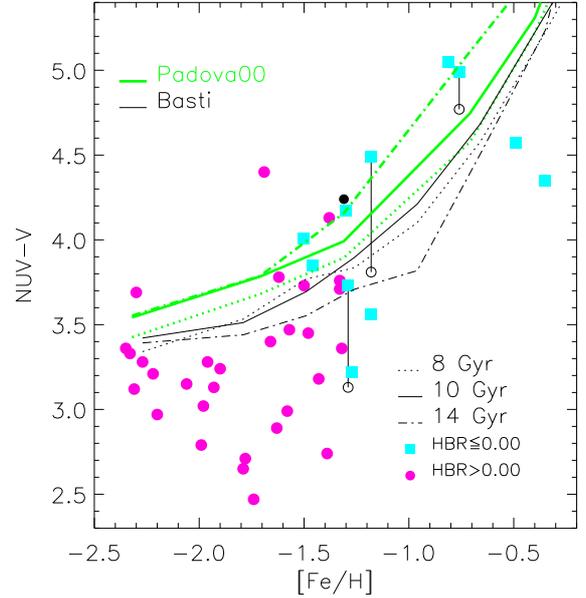}
   \caption{Dependence of the integrated NUV-V colour on metallicity for
Galactic globular clusters. Symbols represent the GALEX data from
\citet{Dalessandro12}. Magenta circles (cyan squares) represent data with blue (red)
horizontal branch morphology, according to the HBR parameter from Harris
(2010).   Open circles represent the integrated colours measured from aperture
photometry for the three clusters with large deviation from the surface profile
fitting method. Overplotted are our model predictions based on the Padova00
(thick green
lines) and BaSTI isochrones (thin black lines) for three SSP ages, as indicated. Old
SSP models based on BaSTI provide a reasonably good fit to the data for metallicities above
$\Feh=-1.5$\,dex.}
   \label{fig:GC}%
\end{figure}

In Fig. \ref{fig:GC} we show the integrated GALEX colours of the Galactic
globular clusters from \citet{Dalessandro12}. The integrated magnitudes are
measured from the surface brightness profile fitting method, that for most of
the clusters agrees well with the aperture photometry method. The three
clusters showing large discrepancies among the two methods are shown as open
circles in Fig. \ref{fig:GC}.  Our models are shown for ages ranging from $8$
to $14$\,Gyr and Kroupa Universal IMF. The Padova-based models show a fair
agreement with data for nearly solar metallicity, but they appear too red in
comparison to the metal-poor clusters. In the case of BaSTI-based models, the
match with metal-poor clusters slightly improves. Note that due to the 
differences in the core He-burning phase (see Section\,\ref{sec:models}), the
metal-poor BaSTI-based models become bluer at old ages (see Fig.
\ref{fig:NUVV}). Thus the oldest SSP ($14$\,Gyr) is the one providing the
best-fit to the data. Note also that at metallicity below $-1.7$ our models
are not safe and the comparison with data should be taken with caution.

To understand whether the HB morphology can have an impact on the integrated
colours, we colour-code the clusters according to their HBR parameter (from
\citealt{Harris10}). Clusters with blue and red morphology segregate in the
high and low metallicity part of the diagram, respectively. For the clusters
with $\Feh=-1.3$, there is a hint suggesting that our models match better the
redder HB morphologies, but, given the dispersion of the data, we cannot
conclude that there is clear dependence of the NUV$-$V colour on the HBR
parameter \footnote{We have also looked for the Blue Stragglers (BS)
determination in this sample of clusters and found 25 clusters with BS
determination from \citet*{Moretti08} but no clear impact of the presence of
BS on the colours is observed}.

\subsubsection{Galaxies}
\label{sec:ETGscolours}

\begin{figure}
   \centering
  \includegraphics[width=0.49\textwidth]{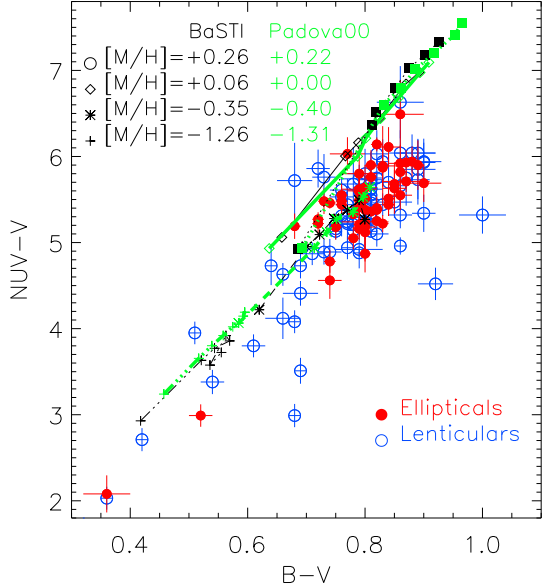}

   \caption{NUV$-$V vs. B$-$V colour-colour diagram for the nearby galaxy
sample of \citet{Donas07}. The colours, which are given in AB magnitudes. are
shown for galaxies classified as ellipticals (red solid circles) and
lenticulars (blue open circles). Overplotted are our model predictions for SSP
models based on Padova00 (green thick lines) and BaSTI (black thin lines) isochrones, for
a Kroupa Universal IMF and different metallicities, as indicated by different
line-types and quoted within the panel. The lines join SSP models of the same
metallicity and ages ranging from $2$ to $14$\,Gyr, from the bottom-left to
the top-right. The plotted ages are $2, 5, 6.3, 8, 10, 12.5$ and $14$\,Gyr for
the predictions based on the Padova00 isochrones and $2, 5, 6, 8, 10, 12$ and
$14$\,Gyr for the models based on BaSTI isochrones. The model colours have
been obtained after redshifting the SSP spectra to $z=0.01$, i.e. the median
redshift of the galaxy sample.}

   \label{fig:GAL1}%
\end{figure}

In Fig.\,\ref{fig:GAL1} we compare our  model predictions with data from
nearby galaxies with GALEX observations, taken from \citet{Donas07}. To
compute the synthetic colours, our models have been redshifted to $z=0.01$,
i.e. the median redshift of the observational sample. Both the ultraviolet
and the optical colour (B$-$V) are given in the AB system and the
observational data are corrected for Galactic extinction. In
Fig.\,\ref{fig:GAL1} we show the comparison with SSP models. Our models at
subsolar and very low metallicity fit the galaxies with blue and intermediate
colours. However the old models of solar and supersolar metallicity are too
red to cover the redder, more massive, subsample of galaxies. 

This mismatch cannot be attributed to the effects of the IMF, which is claimed
to be bottom-heavy for such massive objects \citep{LaBarbera13,Spiniello14}.
These blue colours show little sensitivity to this parameter as shown in
Fig.\,\ref{fig:NUVVimf}. It can be also argued that our NGSL based models
follow the chemical evolution pattern of the solar neighborhood, being
scaled-solar around the solar metallicity regime, whereas massive galaxies are
nevertheless known to be \MgFe-enhanced \citep*{Worthey92}. Therefore
self-consistent SSP spectra with varying \MgFe\ ratios in this spectral range
are required to test this possibility. In fact, we showed that such models
improve the fits of the optical colours of these galaxies
\citep{MIUSCATII,Vazdekis15}. Although we are unable to show whether a better
agreement is reached, we note that any solution should also match the
line-strengths in this spectral range as we will show in Section
\ref{sec:ETGslines}. Finally, it might be also considered that the UV upturn
phenomenon (e.g., \citealt{CodeWelch79,Burstein88}), which affects the FUV
spectral range of a fraction of these galaxies (e.g.,
\citealt{Kaviraj07,Yi11}), is also affecting the colours discussed here (e.g.,
\citealt{Ponder98}). We have identified a number of well known galaxies
showing the UV upturn (e.g., NGC\,4552, one of the strongest UV upturn passive
galaxies in the nearby Universe) as well as UV-weak galaxies (e.g., M\,32).
Although, for clarity, we do not mark them in Fig.\,\ref{fig:GAL1}, we find
that these galaxies do not deviate significantly from the region where most
galaxies are located. Specifically, the identified UV upturn galaxies do not
prefer the locus of the galaxy branch that cannot be matched by the SSP models
alone.

\begin{figure}
   \centering
  \includegraphics[width=0.49\textwidth]{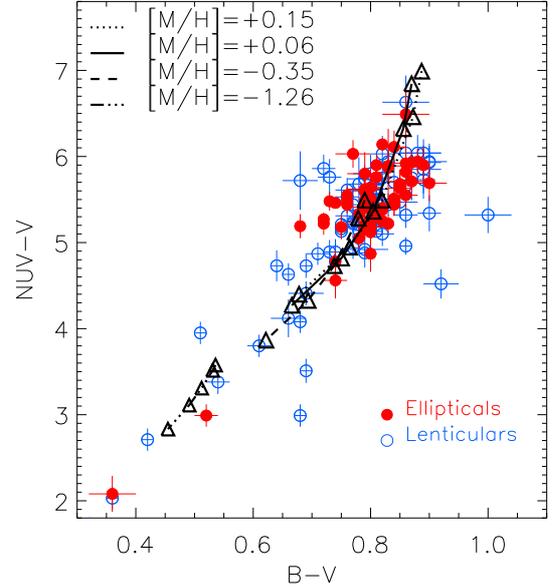}

   \caption{The same galaxy sample as in Fig.\,\ref{fig:GAL1} is plotted here.
Overplotted are model predictions (open triangles) for different metallicities
obtained by combining two SSPs (based on BaSTI and adopting a Kroupa Universal
IMF), one old with age $12$\,Gyr and the other young with $0.5$\,Gyr. The young
component has varying mass fraction, namely $0, 0.1, 0.5, 1$ and $2$\,\% from
top to bottom, respectively.}

   \label{fig:GAL2met}%
\end{figure}
\begin{figure}
   \centering
  \includegraphics[width=0.49\textwidth]{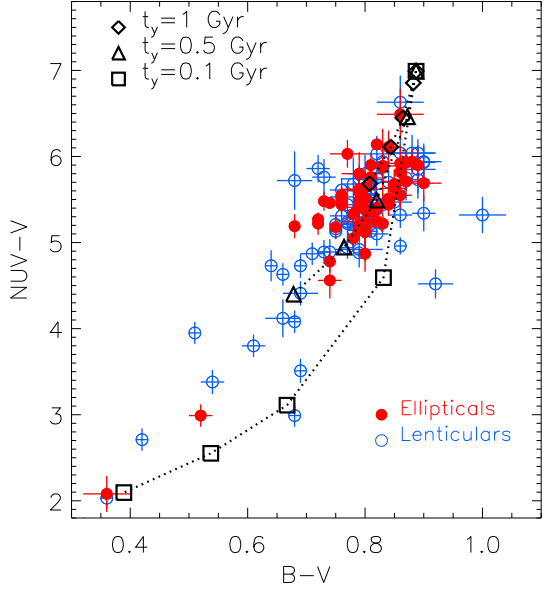}

   \caption{The same galaxy sample as in Fig.\,\ref{fig:GAL1} and
Fig.\,\ref{fig:GAL2met}. Here we overplot combined models with  metallicity
$\Feh=+0.15$, composed of an old SSP of $12$\,Gyr with varying mass fractions of
a young component of $0.1$\,Gyr (open squares), $0.5$\,Gyr (open triangles) and
$1$\,Gyr (open diamonds). The plotted fractions are $0$, $0.1$, $0.5$, $1$ and
$2$\,\%, from top to bottom, respectively.}

   \label{fig:GAL2young}%
\end{figure}

In Fig.\,\ref{fig:GAL2met}, we show the effect of an extended star formation
history on this colour-colour diagram.  We consider two burst models, where
on top of the main burst with age 12\,Gyr we add another burst of $0.5$\,Gyr,
with varying mass fractions, namely $0$, $0.1$, $0.5$, $1$ and $2$\,\%. These
combinations are made for varying metallicities from $\Feh=+0.15$ down to
$\Feh=-1.26$, as quoted within the figure. The addition of the young burst
contributing with less than 1\,\% to the mass in the solar and supersolar regime
is very effective at bluening the colours by $\sim$1\,mag in the NUV$-$V and by
$\sim0.05$ in the B$-$V, thus bringing the models in better agreement with data.
We also note that there is a bunch of galaxies with red B$-$V colour, likely the
most massive ones, which are not properly matched by such combination of models.
For this reason we plot in Fig.\,\ref{fig:GAL2young} other models where we
combine an old component of 12\,Gyr with varying fractions, from $0$ to $2$\,\%,
of a young component of varying age, namely $0.1$, $0.5$ and $1$\,Gyr. Both
components, i.e. the old and the young, have the same metallicity \Feh=+0.15, as
this metallicity might be representative for the most massive galaxies as
derived from detailed spectroscopic studies (e.g., \citealt{LaBarbera13}). We
see that most of the galaxies, including those not matched by the SSP models in
Fig.\,\ref{fig:GAL1} and the combined models in Fig.\,\ref{fig:GAL2met}, are now
well reproduced by a varying fraction/age of the young component. We find that
the best model combinations include young bursts in proportions that range from
$0.1$\,\% of $0.1$\,Gyr to $2$\,\% of $1$\,Gyr, thus explaining the observed
scatter in the galaxy distribution. These results are in very good agreement
with previously reported fractions and ages for the young components (e.g.,
\citealt{Yi11}).

\subsection{Spectra}
\label{ sec:ETGspecra}

\begin{table}
\centering
\caption{\label{tab:toloba} SSP-equivalent ages and metallicities of
a sample of ETGs with varying mass derived from full spectrum-fitting.}
\begin{tabular}{lccc}
Galaxy & $\sigma$ (Km\,s$^{-1}$)$^{a}$ & Age (Gyr) & Metallicity (dex) \\
\hline                   
NGC\,221  & 83 & 3.4$\pm$0.1 & +0.07$\pm$ 0.01 \\
NGC\,3377 &142 & 9.0$\pm$0.5 & -0.07$\pm$ 0.04 \\
NGC\,6703 &191 & 3.6$\pm$0.5 & +0.13$\pm$ 0.05 \\
NGC\,3379 &228 &10.0$\pm$0.3 & -0.24$\pm$ 0.02 \\
NGC\,1700 &253 & 6.2$\pm$0.4 & -0.11$\pm$ 0.03 \\
NGC\,4472 &310 & 8.5$\pm$4.1 & +0.00$\pm$ 0.31 \\
\hline
\\
\multicolumn{3}{l}{$^a$ Taken from \citet{Toloba09}}.
\end{tabular}
\label{tab:tolobafits}
\end{table}

\begin{figure*}
   \centering
  \includegraphics[width=0.9\textwidth]{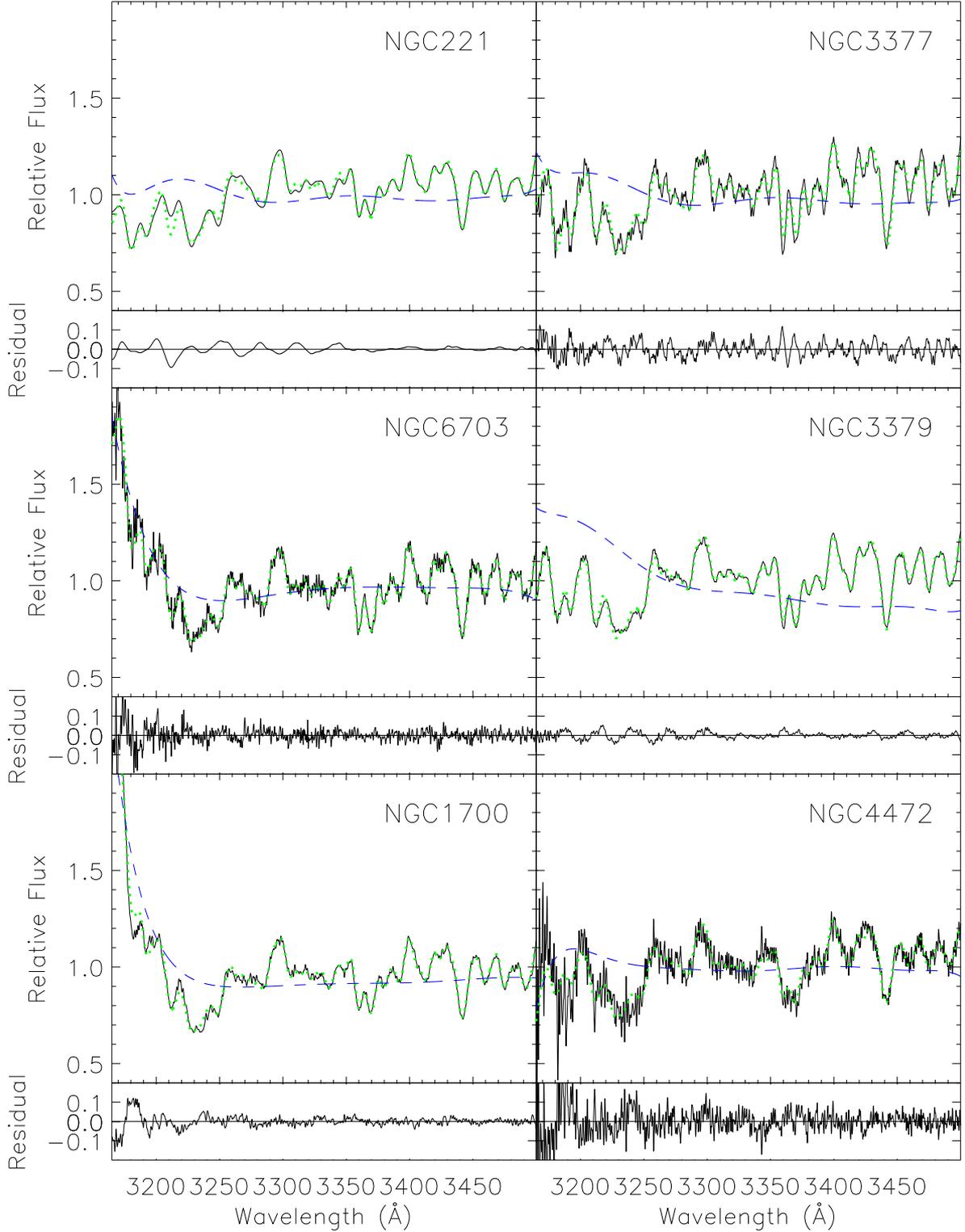}

   \caption{ULySS full spectrum-fits of a representative sample of galaxy
spectra (quoted within the panels) of \citet{Toloba09}. In the upper panel for
each galaxy we plot the observed spectrum in black and the best fitted model
in thick dotted light green. The multiplicative polynomial employed in the fitting is shown in
dashed blue. In the lower panel we plot the residuals (model-observation) with the
same scale used for the spectra. We organized these plots such that the left
panels correspond to galaxies for which we obtained relatively young ages in
comparison to the ones in the right, which are old. The galaxies increase
their velocity dispersion from top to bottom. The ages and metallicities of
the best fitting models are listed in Table\,\ref{tab:toloba}. In the text we
discuss how these estimates compare to those from the literature that are
based on detailed line-strength studies with high quality optical spectra.}

   \label{fig:tolobafits}%
\end{figure*}

\begin{figure}
   \centering
  \includegraphics[width=0.49\textwidth]{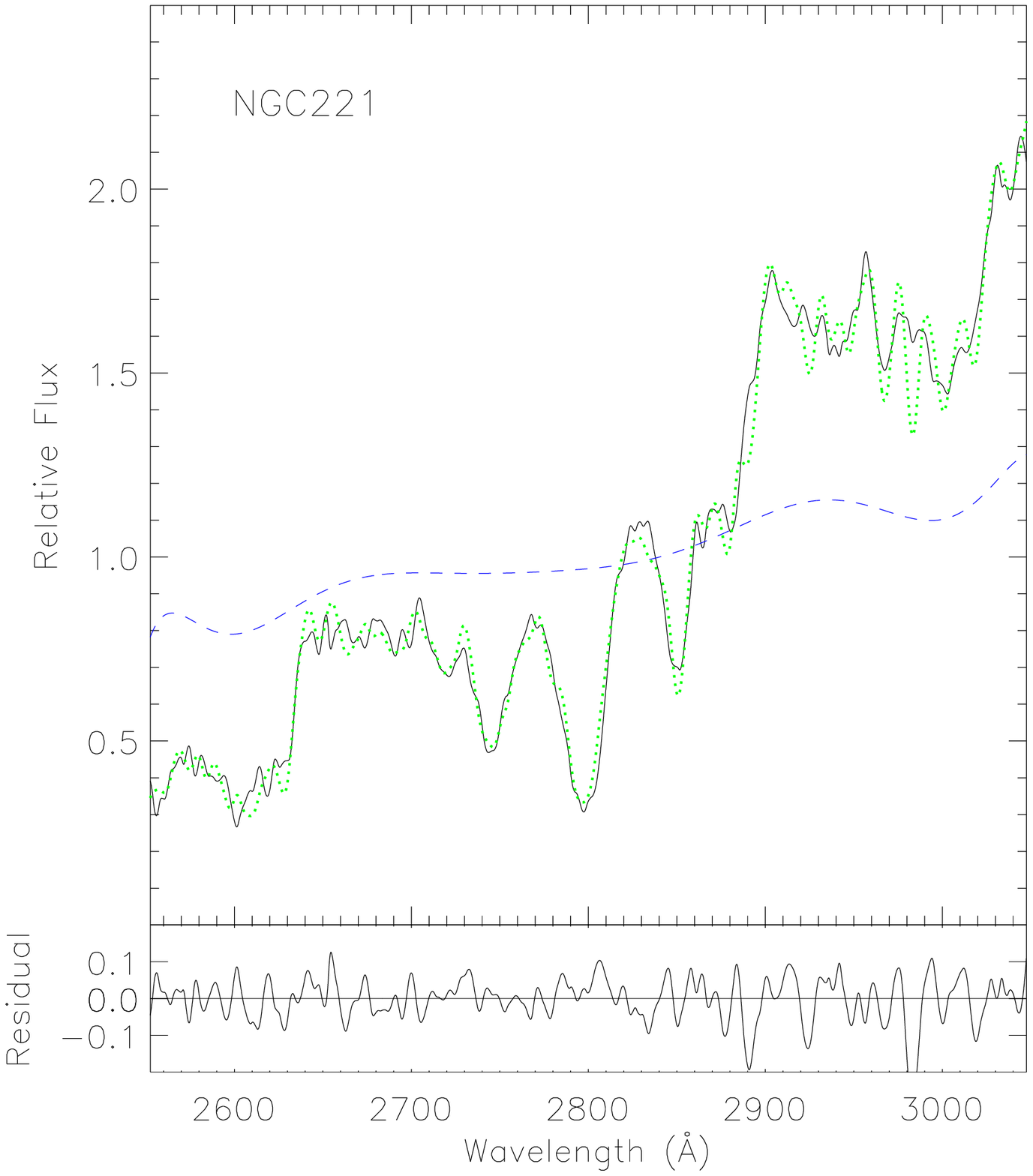}

   \caption{The upper panel shows the ULySS fit of M\,32 in the spectral
region around the Mg features at $\sim2800$\,\AA. This spectrum, observed with
the IUE, from the catalog of \citet*{McQuade95} and \citet*{Storchi-Bergmann95},
is shown in black. Overplotted in thick dotted light green is a model with age $2.3$\,Gyr and
metallicity \Mh$=+0.11$. All the plotted spectra have a resolution of
$\sigma=270$\,\kms. The multiplicative polynomial employed in the fitting is
shown in dashed blue. In the lower panel we plot the residuals (model-observation) with
the same scale used for the spectrum. }

   \label{fig:m32fits}%
\end{figure}

In this section we apply full spectrum-fitting to a number of well selected
ETGs from the sample of \citet{Toloba09}. The galaxies cover a wide mass range
(velocity dispersion between $80$\,\kms\ and $350$\,\kms) and have varying
SSP-equivalent ages, as estimated from the optical spectral range.  Although
the wavelength coverage of the data ranges from $3150$\,\AA\ to $4040$\,\AA,
here we focus on fitting the spectral region blueward the MILES range, i.e.
$\lambda\lambda$ $3150$--$3500$\,\AA. This allows us to show to what extent
our results agree with those obtained in the optical range. 

We use ULySS\footnote{\url{ulyss.univ-lyon1.fr}}\citep{Koleva09}, which is a
full spectrum fitting package that performs $\chi^2$-fit between the observed
spectrum and a linear combination of models. In this case we reconstruct the
observations as SSPs convolved with the line-of-sight velocity dispersion
(LOSVD) of the data, multiplied by a polynomial. We employ an interpolated
grid of SSP spectra computed with Kroupa universal IMF. We injected the line
spread function of the observations into the models to match their resolution
and we used multiplicative polynomial of a degree $8$ to minimise the
mismatch between the high order changes in the observational flux (due to
extinction and instrument) and the models. To fit the LOSVD of the galaxies
with the lowest velocity dispersion (NGC\,221 and NGC\,3377) we smoothed them
with a gaussian kernel of $150$\,\kms.  Finally, we used the automatic
kappa-sigma clipping algorithm in ULySS for the outliers. 

In Fig.\,\ref{fig:tolobafits} we show the fits obtained for six galaxies,
namely, NGC\,221, NGC\,1700, NGC\,3377, NGC\,3379, NGC\,4472 and NGC\,6703.
The galaxies have been ordered according to increasing velocity dispersion
from top to bottom. Galaxies for which the fits indicate very old stellar
populations are plotted on the right panels, whereas their relatively younger
counterparts are shown in the left panels. The obtained age and metallicity
values are listed in Table\,\ref{tab:tolobafits}. Our results for NGC\,221
(M\,32) match those obtained by \citet{SB06b} on the basis of optical
line-strength measurements. Similar results are also obtained by
\citealt{VazdekisArimoto99,Worthey04,Rose05}). We also discuss the NUV indices
of this galaxy in Section\,\ref{sec:ETGslines}). The age that we obtain for
NGC\,3377 is larger than the value reported by \citet{SB06b} ($5$\,Gyr),
although it is similar to their age estimate based on spectral synthesis
technique ($9$\,Gyr). The age obtained for NGC\,6703 is smaller than that of
\citet{SB06b} ($5.5$\,Gyr). For NGC\,3379 these authors obtain a somewhat
smaller age value ($8$\,Gyr). The age obtained for NGC\,1700 is in good
agreement with \citet{SB06b}. The results for this galaxy also agree with the
detailed study performed by \citet*{Kleineberg11}. In this work it is shown
that this galaxy has a counterrotating, kinematically decoupled, core that is
markedly younger than the main body of the galaxy. Finally we obtain for the
cD galaxy in Virgo, NGC\,4472, a good agreement with \citet{SB06b}
($9.5$\,Gyr).

It is remarkable that the fits of such a narrow wavelength range in the NUV 
provide us with age values that are so similar to those obtained from detailed
line-strength studies in the optical range. We therefore conclude that for
these ETGs the dominant populations in this spectral window are not
significantly different from those contributing in the visible.

In Fig.\,\ref{fig:m32fits} we show the ULySS fit obtained for M\,32 in the
spectral range covering the Mg features at $\sim2800$\,\AA. We used for this
purpose the spectrum observed with the IUE and presented in the catalog of
\citet*{McQuade95} and
\citet*{Storchi-Bergmann95}\footnote{\url{http://www.stsci.edu/ftp/catalogs/nearby_gal/sed.html}}.
We find a solution that is $\sim1$\,Gyr younger than that obtained in the redder
wavelength range covered by Fig.\,\ref{fig:tolobafits} or in the optical range.
Although this spectrum, which is among the ones with the highest S/N in the
catalog, has a rather modest quality for this type of analysis, this result
suggests a somewhat more extended SFH than just an SSP of $\sim3$\,Gyr. Note
that this is expected as the obtained solution is biased toward a younger age
when fitting bluer wavelengths. 

\subsection{Line-strengths}

\subsubsection{Stellar clusters}
\label{sec:GCslines}

\begin{figure}
   \centering
  \includegraphics[width=0.49\textwidth]{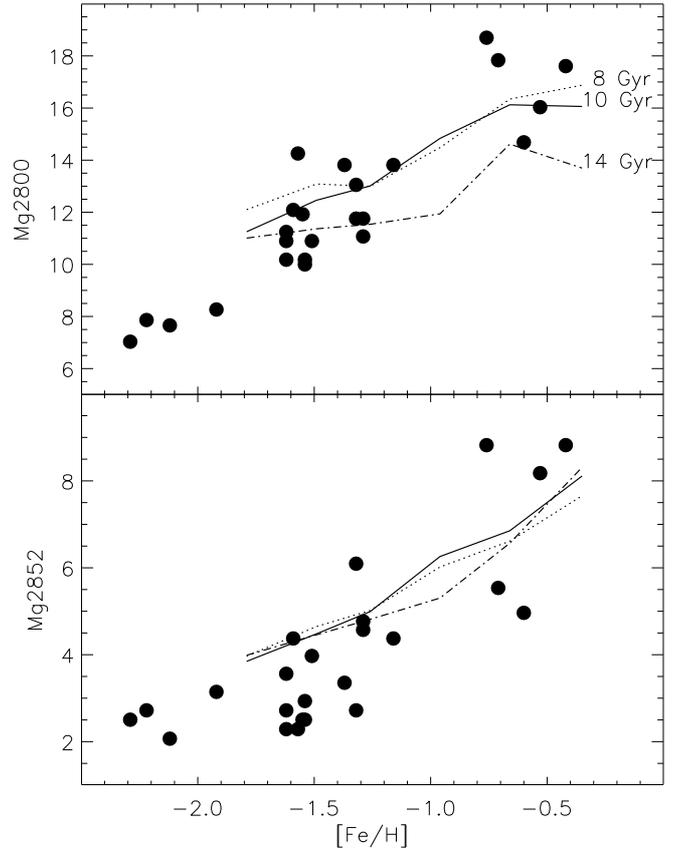}

   \caption{Mg\,2800 (upper panel) and Mg\,2852 (lower panel) UV indices of the
   globular clusters of the Milky Way, observed with the IUE and measured by
   \citet{RoseDeng99}. The metallicities tabulated in the  \citet{Harris96}
   catalog are adopted. Overplotted are our SSP models, based on the BaSTI
   isochrones and adopting a Kroupa Universal IMF, smoothed to match the IUE
   resolution.  The predicted indices for three ages, namely $8, 10$ and $14$\,Gyr,
   are represented with different line types, as quoted within the upper panel,
   as a function of metallicity.}

   \label{fig:GClines}%
\end{figure}

To assess the reliability of our new models we compare the predicted
line-strengths with those of globular clusters, for which Colour-Magnitude
Diagram (CMD) derived ages and metallicities are available. These objects can
be considered to a good approximation a single-burst like stellar population,
which is characterized by a single age and a single metallicity. Therefore
they can be treated as real SSP templates to test our model SSPs, with the
caveats that many clusters seem to be composed by more than just one stellar
component (e.g., \citealt{Meylan03,CassisiSalaris13}) and that stochastic
effects may be noticed in their integrated spectra (e.g.,
\citealt{Cervino02}). We also note that the claimed difference in age among
the various stellar components within a cluster is certainly negligible in
comparison to their old ages, while the difference in the abundance ratios of
individual elements might be more notorious in some line indices (e.g.,
\citealt*{Piotto12,Gratton12}). 

We use the line-strength measurements of \citet{RoseDeng99} for the Milky Way
clusters, which were observed with the IUE. No errorbars are provided by these
authors. Fig.\,\ref{fig:GClines} shows the observed values for the Mg\,2800
and Mg\,2852 indices, as well as those of our SSP models once smoothed to
match the IUE resolution. We show the predictions for three age values ($8$,
$10$ and $14$\,Gyr), which are representative of the globular cluster ages, as
a function of metallicity. The metallicities of the clusters are taken from
\citet{Harris96}. Overall we see that our models match reasonably well the two
indices. For the Mg\,2800 index there is a hint of a slightly younger age for
the metal-rich clusters, although this is not conclusive given the large
scatter of the data. However most of the metal-poor clusters tend to be bluer
than our predictions, similarly to the results obtained for the NUV$-$V
colour, which can be explained by a bluer HB morphology for these clusters
than is adopted by the models. 

\begin{figure}
   \centering
  \includegraphics[width=0.49\textwidth]{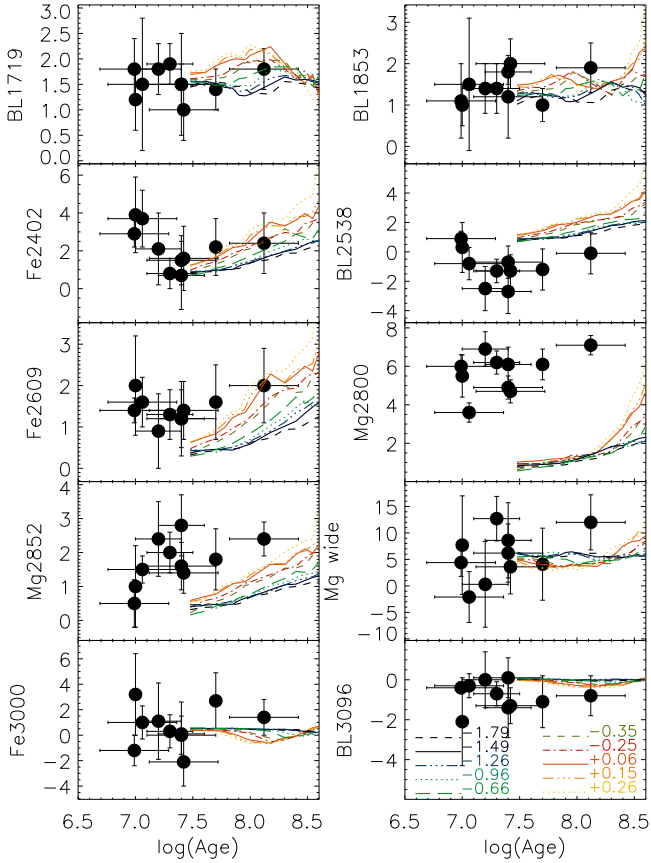}
   \caption{UV line indices for the LMC clusters plotted against their
   CMD-derived ages, all taken from \citet{Maraston09}. Our young BaSTI-based
   SSP models with Kroupa Universal IMF, smoothed to match the IUE resolution, 
   are plotted for different metallicities
   with different line types and colours, as quoted within the last panel.} 
   \label{fig:LMClines}%
\end{figure}

To validate our youngest models we employ observations of the Large Magellanic
Cloud (LMC) stellar clusters. In Fig.\,\ref{fig:LMClines} we compare our
models to the line-strength measurements of \citet{Maraston09} using IUE data
from \citet*{Cassatella87}, which are plotted as a function of the CMD ages
derived by various authors \citep{ElsonFall88,Elson91,Dirsch00,deGrijs02}. A
comparison of this figure with figures 9 and 10 of \citet{Maraston09} shows a
good agreement between our predictions and their models based on the IUE
stellar spectra. The main differences are generally found for ages above
$100$\,Myr. Also, for the BL\,3096 index our model values are about
$0.5$\,\AA\ smaller than theirs, in better agreement with the data. Finally we
also note that a worse agreement is reached when compared to their fully
theoretical model predictions.

Most of the LMC clusters plotted in Fig.\,\ref{fig:LMClines} are younger than
our models with the lowest age, i.e. $30$\,Myr ($\log({\rm Age})=7.48$). However
we see that an extrapolation of the model lines toward younger ages match most
of the plotted indices within their errorbars. This is the case for the
BL\,1719, BL\,1853, Fe\,2402, Mg wide, Fe\,3000 and BL\,3096 indices. Our models
provide slightly larger values for the BL\,2538 and slightly smaller index
strengths for Fe\,2609 and Mg\,2852 indices. However a strong discrepancy is
found for the Mg\,2800 index, for which our models provide around $5$\,\AA\
smaller values. Note that this is remarkable, given the good agreement found for
the Milky Way old clusters, as shown in Fig.\,\ref{fig:GClines} and our good
match to the broadband Mg wide index. A similar discrepancy is obtained when
these clusters are compared to the \citet{Maraston09} models (see their Figure
10). These authors attribute this strong Mg absorption to contamination by
interstellar lines from warm neutral interstellar medium of the LMC, which
affects this Mg region. We refer the interested reader to \citet{Maraston09} for
a more extended discussion on this discrepancy.

\begin{figure*}
   \centering
  \includegraphics[width=0.85\textwidth]{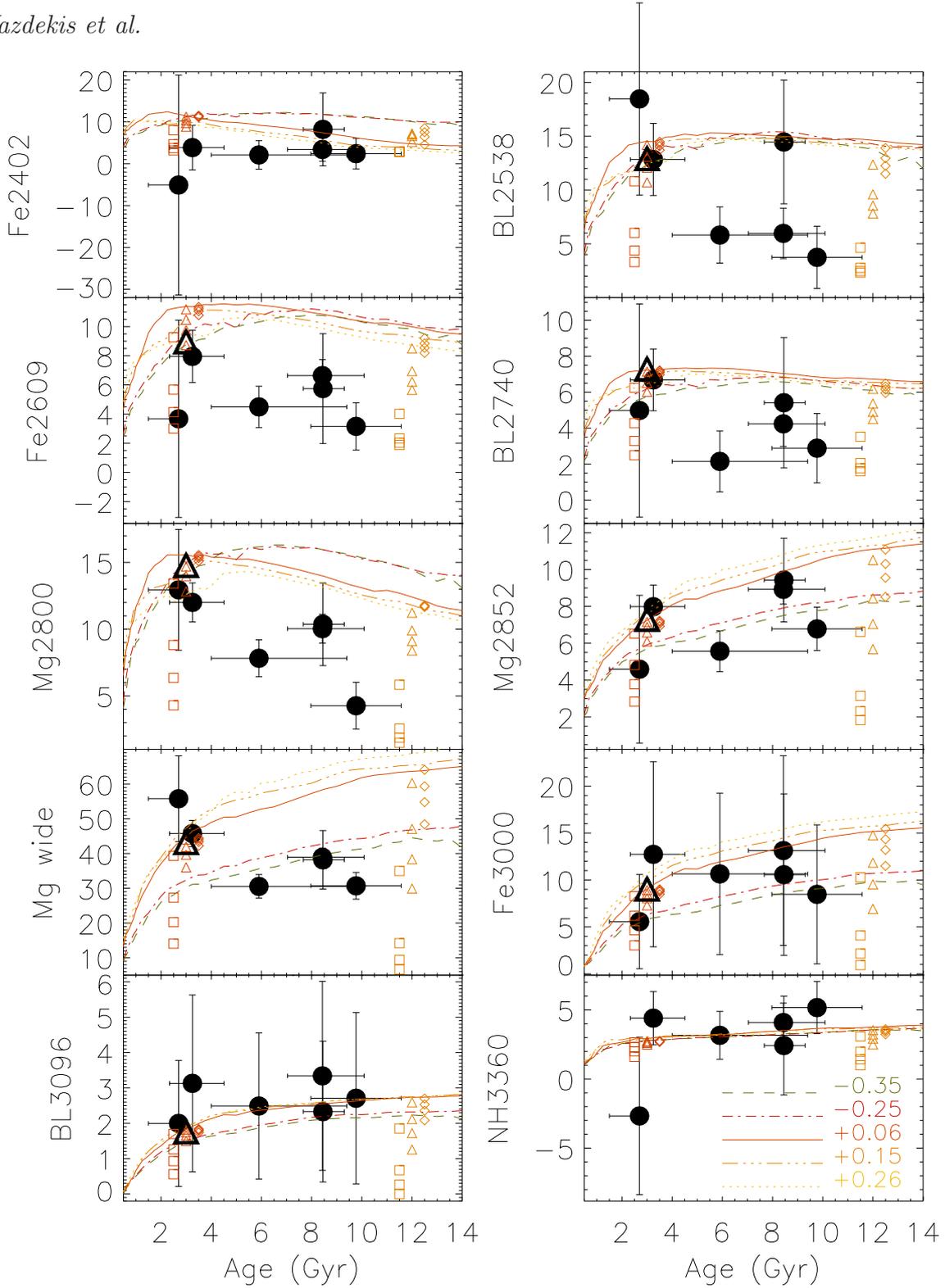}
   \caption{UV line-strengths of elliptical galaxies in the \citet{Ponder98}
sample.  The galaxies, with varying velocity dispersion, were observed with FOS
at the HST, and are shown as filled circles. M\,32, which was observed with the
IUE, is plotted as an open triangle. These galaxies display a rather wide
coverage in mean luminosity-weighted age as estimated from the optical spectral
range (see the text for references). Our SSP model predictions for different
metallicities are shown in coloured lines and varying types as quoted within the
last panel. Both galaxy and model line-strengths are shown at a resolution of
$300$\kms. We also show the line-strengths resulting from combining a $12$\,Gyr
model with a varying fraction of a young model with age $0.1$\,Gyr (open orange
squares), $0.5$\,Gyr (open orange triangles) and $1$\,Gyr  (open orange
diamonds), all with supersolar metallicity $\Feh=+0.15$, which can be considered
representative for the oldest (massive) galaxies.  The mass fraction of the
young component is increased by $0.1$, $0.5$, $1$ and $2$\,\% from top to
bottom, resulting in an increasing departure from the reference SSP  value at
$12$\,Gyr, which is located on the model line (three dots - dashed)
corresponding to this metallicity at $12$\,Gyr.  These symbols were separated
arbitrarily from the reference age value for visibility. Similarly, we also plot
the combination of a model of age $3$\,Gyr with young models with similar ages
and fractions, all with metallicity $\Feh=+0.06$. We use similar symbols, this
time in red and placed at  $3$\,Gyr. The larger the young mass fraction the
larger the deviation from the reference SSP value located on the solid red line
at this age.}
   \label{fig:ETGslines}%
\end{figure*}

\subsubsection{Early-Type galaxies}
\label{sec:ETGslines}%

To compare our predicted line-strengths to index measurements of ETGs we
selected the sample of six elliptical galaxies of \citet{Ponder98} (NGC\,3605,
3608, 5018, 5831, 6127 and 7619). The galaxy spectra were obtained with the
Faint Object Spectrograph on the HST covering the UV spectral range at a
resolution of $8$\,\AA, with S/N$>20$ and flux accuracy of $\sim5$\,\%.
Fig.\,\ref{fig:ETGslines} shows these galaxies (solid circles), with their
errorbars, for $10$ UV line-strength indices. We assume for these galaxies the
mean luminosity-weighted ages derived from the optical range based on
line-strength studies.  We adopt the ages of \citet{SB06b} for NGC\,3605, 3608
and 6127 ($3.25$, $8.43$ and $8.45$\,Gyr, respectively),
\citet{LeonardiWorthey00} for NGC\,5018 ($2.8$\,Gyr), with an errorbar adjusted
to match the younger age value provided by \citet{TerlevichForbes02} (i.e.
$1.5$\,Gyr), \citet{Vazdekis04} for NGC\,5831 ($5.9$\,Gyr), and
\citet{GonzalezDelgado15} for NGC\,7619 ($9.77$\,Gyr). In addition we also
include the index values provided by \citet{Ponder98} for M\,32, obtained with
the IUE by \citet*{Buson90}. No errorbars are provided for this galaxy. We adopt
the central age value obtained by \citet{Rose05} ($3$\,Gyr). The galaxies cover
a rather wide range in velocity dispersion, from $\sigma\sim80$\,\kms\ (M\,32)
to $\sim350$\,\kms\ (NGC\,7619). Note that we did not attempt to homogenize
these age estimates, which may vary among authors employing different age-dating
line-strength based methods and models. We selected studies based on spectra of
very high quality, which suffices for the purposes of this comparison. Roughly,
the age of these galaxies increases with increasing $\sigma$, with the notably
exception of NGC\,5018 ($280$\,\kms), our youngest object, which is considered a
remnant merger \citep{Fort86}.

It is worth quoting that these old galaxies show Mg\,2800 values significantly
smaller than those shown by the MW globular clusters as seen in
Fig.\,\ref{fig:GClines} (regardless a relatively minor correction due to the
varying velocity dispersion), in good agreement with the expected trend with
varying metallicity from our model predictions. However we see that the
observed indices for these massive galaxies are smaller than those predicted
by our metal-rich SSP models. In fact, in contrast to M\,32, whose measured
indices are well matched by a solar metallicity SSP with age in good agreement
with its literature determination ($\sim3$\,Gyr), this is not the case for
various of the old massive galaxies. Specifically our metal-rich SSP models
predict larger values for the BL\,2538, Fe\,2609, BL\,2740, Mg\,2800,
Mg\,2852, Mg wide and Fe\,3000 indices. 

A bottom-heavy IMF, as recently claimed for massive early-type galaxies (e.g.,
\citealt{LaBarbera13, Spiniello14}), decreases the Mg\,2800 index by only
$\sim1$\,\AA\ for an SSP with a bimodal IMF with slope $2.8$, which is clearly
insufficient for matching the observed index values of these massive objects.
Small index variations are also obtained for the other indices, confirming the
very small sensitivity of this spectral range to the IMF, as already found for
the colours (see Fig.\,\ref{fig:NUVVimf}).

We also investigate whether the existence of an old metal-poor contribution in
addition to a dominant metal-rich population (e.g.,
\citealt{Vazdekis97,MarastonThomas00}) is capable of matching the observed index
values. Note however that the observed values deviate in different proportions
and sometimes in different direction to the expected index variation due to a
decrease in metallicity. Therefore, if we construct a
model that is composed of two old SSPs, one metal-rich and the other one
metal-poor, we could in principle match the Mg wide index. On the contrary we
are unable to match simultaneously the Mg\,2800 index, for which we will predict
a larger value in comparison to a metal-rich SSP. It can be argued that the
Mg\,2800 could be affected by chromospheric emission filling-in, thus making the
index smaller. However such a model combination cannot match for example the
Fe\,2609 and Fe\,3000 pair of indices simultaneously either. Therefore we can
safely rule out such a combined model for matching the UV line-strength indices
of these massive galaxies.

As an alternative scenario we also study the impact of a young component on
these indices. Such a scenario has been proposed in the literature to explain
for example the scatter of the Balmer absorption lines in the optical range
(e.g., \citealt{LeonardiRose96,Trager00}). For this purpose we plot in
Fig.\,\ref{fig:ETGslines} the line-strength values resulting from combining an
model with age $12$\,Gyr with a young model with age $0.1$, $0.5$ and
$1$\,Gyr, all with metallicity $\Feh=+0.15$, which is representative for these
massive galaxies. The mass fraction of the young component is varied from
$0.1$ to $2$\,\%, which is reflected by an increasing deviation from the
reference SSP value, which is located on the model line of the same
metallicity at $12$\,Gyr. We see that all the indices are matched for
the model combination including 0.1\,\% of a young component of age
$0.1$\,Gyr. Also, a reasonably good match is achieved when we include
$0.5$--$1$\,\% of a population with age $0.5$\,Gyr. This set of UV indices is
not only allowing us to detect young contributions but also to constrain their
ages. In fact it can be seen that when using a component of $1$\,Gyr there are
indices that cannot be matched such as BL\,2538, BL\,2609, BL\,2740 or
Mg\,2800. Such a combination can be therefore ruled out. Note that what really
matters is the age of this component, as the impact of varying the fraction is
much smaller. Therefore these indices allow us to break the burst-age,
burst-strength degeneracy \citep{LeonardiRose96}. Note that the effects of
these young contributions is far less notorious for the reddest indices, i.e.
as we approach to the optical range. Also the effects are relatively small for
our bluest index Fe\,2402, in this case because these young populations still
show as large index values as the old component.  

It is worth noticing that our purpose with these model combinations is not to
properly fit these data, but to illustrate the abilities of this set of UV
indices to constrain the presence of such young stellar components. The ages and
fractions obtained with these indices are in good agreement with those inferred
from the colours as shown by Fig.\,\ref{fig:GAL2young}. In principle, we also could
explain the observed index scatter shown by these old galaxies by a varying age,
and to a lesser extent by a varying fraction, of these young components.

It is remarkable the very good agreement achieved for M\,32, with just an SSP of
age similar to that corresponding to the mean luminosity-weighted age derived
from the optical range. The center of this galaxy is known to be a factor of
$\sim$2 younger than at $1$\,\re\ (e.g., \citealt{Worthey04,Rose05}). Such a
relatively young mean luminosity-weighted age in the optical range suggests the
presence of a population with age smaller than the determined mean optical age
value. It is therefore expected that such a population is largely dominating the
light of this galaxy in the UV range. In fact, the younger the mean optical age
the more relevant is the contribution of this relatively young population, and
far smaller that of the  old component.  In Fig.\,\ref{fig:ETGslines} we also
show what impact the presence of even younger populations has on a dominating
population with age $3$\,Gyr and metallicity $\Feh=+0.06$. We use the same ages
and fractions employed to illustrate the effects on a dominant old population.
We see that the impact of these younger contributions is notoriously less
influential on the indices than it was the case for older galaxies. We note
specifically that the ages and fractions of the young stellar components that
are required to match the UV indices for the old galaxies (i.e. $0.1$\,\% of
$0.1$\,Gyr and $0.5-1.0$\,\% of $0.5$\,Gyr) have little impact on the indices of
the dominating population of 3\,Gyr. To create a noticeable effect, larger
fractions and younger contributions are needed, as illustrated by the
combination with a $0.1$\,Gyr and mass fractions $0.5$, $1$ and $2$\,\%. Overall
we also see that these fractions tend to provide a slightly better agreement for
the other young galaxies in this sample. To summarize, the population with age
$\sim3$\,Gyr completely dominates the integrated light of these galaxies in the
UV. Note, however, that our ULySS fit to this galaxy in the spectral range
around the Mg features at $\sim2800$\,\AA\ suggests a somewhat more extended SFH
as we estimate a slightly younger age. A more quantitative study is beyond the
scope of this paper but will be addressed elsewhere. However these results
illustrate the ability of the UV to detect and constrain the contribution of
young populations that do not show up so clearly in redder spectral ranges,
particularly for the old galaxies. 

\section{Summary and Conclusions}
\label{sec:summary}

We employ the New Generation Stellar Library (NGSL), which was observed with the
Hubble Space Telescope, to extend our stellar population synthesis models to the
UV spectral range. The stellar parameters coverage of this library, which were
determined in \citet{NGSLI} and transformed to match those of the MILES library
\citep{MILESII} on which our models in the optical range are based, allowed us
to compute single-age, single-metallicity stellar population (SSP) spectra in
the metallicity range $-1.79<\Mh<+0.26$ and ages larger than $30$\,Myr.
These models represent a significant improvement over earlier studies based on
the IUE stellar library \citep{IUEI,IUEII} as the NGSL feeding our
models has a better stellar parameter coverage, resolution and S/N. 
We provide a quantitative assessment of the quality of these models
within these ranges. The models were computed with two sets of isochrones,
Padova00 \citep{Padova00} and BaSTI \citep{Pietrinferni04, Pietrinferni06} as
extended in \citet{Vazdekis15}. These isochrones were transformed to the
observational plane with metallicity dependent colour-temperature relations
from extensive photometric empirical (rather than theoretical) stellar
libraries (mainly \citealt{Alonso96,Alonso99}). The models were computed for a
suite of IMF types: Kroupa Universal and Revised \citep{Kroupa01}, \citep{Chabrier} and the
single-power law (unimodal) and double-power law low-mass tappered (bimodal)
IMFs of \citep{Vazdekis96}. We also varied the slope of the unimodal IMF and
the high-mass end (M$>0.6$\,\Msol) slope of the bimodal IMF. 

The computed SSP spectra in the UV range cover $\lambda\lambda$
$1680.2$--$3541.4$\,\AA, with constant resolution FWHM$=3$\,\AA\ for
$\lambda<3060.8$\,\AA\ and FWHM$=5$\,\AA\ for $3060.8<\lambda<3541.4$\,\AA. These SSP
spectra were joined to those computed with MILES \citep{MILESI} in the optical range, as well
as with our models for redder wavelengths, all computed with the same code and
prescriptions and employing empirical stellar libraries, i.e. Indo-US
\citep{Valdes04}, CAT \citep{CATI,CATII} and IRTF \citep{IRTFI,IRTFII}, to
obtain self-consistent E-MILES SSP spectra covering the range $\lambda\lambda$ 
$1680$--$50000$\,\AA\ at moderately high resolution (FWHM$=2.5$\,\AA\ from
$3541.4$\,\AA\ to $8950.4$\,\AA\ and $\sigma=60$\,\kms\ for larger wavelengths). As
a sanity check, we compared the U-B colour measured on the E-MILES spectra to
those based on the photometric predictions (using the same photometric libraries
employed to transform the isochrones to the observational plane), reaching a
good agreement (typically within $0.02$\,mag). The reliability of the E-MILES
SSP models depend on the spectral range. We refer the reader to the specific
papers, listed in Table\,\ref{tab:SEDproperties}, of the models joined here for
assessing their quality.

We focus on studying the behaviour of the colours, spectra and line-strength
indices of the SSP spectra in the UV range. The NUV-V colour shows a steady
increase with increasing age (and metallicity), in contrast to the more
flattened behaviour shown by the optical colours for ages above $\sim1$\,Gyr.
Note that such a flattening is far more pronounced for the near-IR colours. We
show that the effects of the IMF on the NUV-V colour is almost negligible. We
also study the behaviour of the UV line-strength indices as a function of
relevant stellar population parameters. We find that some UV indices increase
with increasing age and metallicity, as it happens for most of the metallicity
indicators in the optical range. However, there are other indices within the
range $2300$--$2800$\,\AA, such as Mg at $2800$\,\AA, which peak at
intermediate ages in the range $1$--$6$\,Gyr for metal-rich stellar
populations, whereas for lower metallicities they keep strengthening with
increasing age. The indices within the spectral range $3000$--$3500$\,\AA\
show a larger sensitivity to the age in comparison to their counterparts in
the optical range. 

We compare our models to other models in the literature that employ the IUE
\citep{IUEII}, i.e. an alternative space-based stellar library with lower
spectral resolution. Our NUV-V colour is in reasonably good agreement with the
models of \citet{Maraston05}. It also agrees with those of \citet{BC03},
except for ages above $\sim10$\,Gyr, which are $\sim1$\,mag bluer than ours.
This is attributed to the fact that these models incorporate some spectra of
planetary nebulae that shine in the shortest wavelengths of the spectral range
covered by our SSP spectra. The UV line-strengths of our SSP models with ages
smaller than $\sim0.5$\,Gyr are in good agreement with those shown by
\citet{Maraston09}. For older ages, the peculiar behaviour shown by the
Mg\,$2800$ index as a function of age for metal-rich stellar populations, is
also seen in the models of \citet{BC03}, as shown in \citet{Daddi05}. 

Our models match reasonably well the integrated NUV-V colours of the Milky-Way
globular clusters of higher metallicities as well as a fraction of the
metal-poor clusters. However for other metal-poor clusters our colours are
redder by $\sim0.5$\,mag. Our models provide good fits to the Mg\,2800 and
Mg\,2852 indices of the Milky Way globular clusters. Moreover, most of the UV
line-strengths of the LMC clusters, with ages typically smaller than
$100$\,Myr, are reasonably well fitted by our models. However, a strong
discrepancy is found with respect to the Mg\,2800 measured in these clusters,
confirming the result previously reported by \citet{Maraston09}. 

The comparison of our models with UV data of ETGs, shows the unprecedented
power of this spectral range to constrain the contributions from young stellar
populations to the total light, particularly for the oldest objects. We find
that both, the NUV-V colour and the UV line-strengths, of massive ETGs reveal
the presence of young stellar components on top of the old stellar
populations, with ages in the range $0.1-0.5$\,Gyr that represent $0.1-0.5$\%
in mass fraction. The remarkable variety in the behaviour of the different UV
line indices as a function of age, has allowed us to rule out stellar
contributions with ages around $\sim$1\,Gyr. Our age/fraction estimates for
these young components are in agreement with previous results obtained in this
spectral range (e.g., \citealt{Yi11}). We also find that the age of these
young components could be more important than their relative contributions.
Variations in the ages, and in the relative fractions, of these younger
components might potentially fill-in the scatter observed for the NUV-V colour
and the UV line-strengths of ETGs. It is worth noticing that such tiny
contributions do not affect significantly the age estimates obtained from the
age indicators in the optical range, such as H$\beta$ (these young components
will make these old galaxies to look $1-2$\,Gyr younger if measured with this
Balmer index).

For M\,$32$ we obtain very good fits using an SSP model of $\sim$3\,Gyr (and
solar metallicity). This is attributed to the fact that this galaxy is claimed
to be dominated by a stellar population of such age, which outshines these small
contributions from the young components that we infer for the older, more
massive, ETGs.

We also performed full-spectrum fits in the spectral window $\lambda\lambda$
$3200$--$3500$, using UlySS \citep{Koleva09}, for a representative set of ETGs
with varying mass and ages. The resulting SSP age and metallicity estimates are
in very good agreement with detailed line-strength studies performed in the
optical range. Note that the results obtained for the most massive objects in
our sample, i.e. the same ages as inferred from the optical range, do not
contradict the ones that we obtained by fitting the UV line-strengths, as the
indices that are able to constrain the presence of young contributions are found
at bluer wavelengths ($<3000$\,\AA). Put in other words, the spectral range
employed for our full-spectrum fits provides similar constrains as the optical
range. However our fit of NGC\,221 in the spectral region covering the Mg
features around $2800$\,\AA\ suggests a more extended SF, rather than just a
single $\sim3$\,Gyr old dominating population.

The E-MILES models computed here provide new means for constraining the most
likely Star Formation Histories experienced by massive galaxies. In particular
the extremely wide spectral coverage of these models can be used to separate
the various stellar populations contributing to the total light in the various
wavelength ranges. The ample coverage in age, metallicity and IMF, allow us to
study smaller galaxies and stellar clusters of varying ages. The UV range
covered by these models will help to understand the evolution of these objects
as seen at different redshift ranges. These models, which are aimed for
interpreting incoming data from ground- and space-based observing facilities,
can be retrieved from the MILES website \url{http://miles.iac.es}. The webpage
also provides user-friendly tools for handling the model spectra for their use
for the analysis of the data.

\section*{Acknowledgments} 

We are grateful to M. Beasley, J. Beckman, F. La Barbera and R. Peletier for
very useful discussions and valuable suggestions. We thank G. Bruzual for
providing us with various versions of his models in the UV spectral range and
for interesting suggestions. We also thank E. Toloba for providing us with
fully-reduced early-type galaxy spectra. We are grateful to the anonymous
referee for valuable suggestions that helped us to improve the paper. This
research has made an extensive use of the SIMBAD data base and VizieR catalogue
access tool (both operated at CDS, Strasbourg, France), the NASA's Astrophysics
Data System Article Service. This work has been supported by the grants
AYA2013-48226-C3-1-P and AYA2013-48226-C3-2-P from the Spanish Ministry of
Economy and Competitiveness (MINECO) and the Generalitat Valenciana under grant
PROMETEOII/2014/069. E.R. acknowledges a Marie Heim-V\"ogtlin grant from the
Swiss National Science Foundation. 


\bibliographystyle{mnras}
\bibliography{papers}

\end{document}